\documentclass[aps,twocolumn,prb,showpacs,floatfix,amsmath,amssymb,superscriptaddress]{revtex4-1}
\usepackage{graphicx}
\usepackage{bm}
\usepackage{hyperref}
\usepackage{color}
\usepackage{natbib}
\usepackage{changes,cancel,amsmath, amssymb}
\usepackage{braket,float,soul}
\usepackage{comment}
\usepackage{ulem}

\definecolor{ao(english)}{rgb}{0.0, 0.5, 0.0}

\graphicspath{{./Figures-bs/} } 
\begin{document}
\title{Majorana-like localized spin density without bound states \\in topologically trivial spin-orbit coupled nanowires}



%
\author{Lorenzo Rossi}
\email{lorenzo.rossi@polito.it}
\affiliation{Dipartimento di Scienza Applicata e Tecnologia, Politecnico di Torino, 10129 Torino, Italy}

\author{Fabrizio Dolcini}
\affiliation{Dipartimento di Scienza Applicata e Tecnologia, Politecnico di Torino, 10129 Torino, Italy}

\author{Fausto Rossi}
\affiliation{Dipartimento di Scienza Applicata e Tecnologia, Politecnico di Torino, 10129 Torino, Italy}

\begin{abstract}
In   the topological phase of spin-orbit coupled nanowires Majorana bound states are known to localize at the nanowire edges and to exhibit a spin density  orthogonal to both the magnetic field and the spin-orbit field.  By investigating a nanowire   exposed to a uniform magnetic field  with an interface between regions with different spin-orbit couplings, we find that  the orthogonal spin density is pinned at the interface even when both interface sides are  in the topologically trivial phase,  and even when no bound state is present at all. 
A trivial  bound state may additionally appear at the interface, especially if the spin-orbit coupling takes opposite signs across the interface. However, it can be destroyed by 
 a smoothening of the spin-orbit profile or by a magnetic field component parallel to the spin-orbit field. In contrast, the orthogonal spin density persists in various and realistic parameter ranges. We also show that, while the measurement of bulk equilibrium spin currents has been elusive so far,  such robust orthogonal spin density peak may  provide a way to  detect   spin current variations across interfaces. \\
\end{abstract}

\maketitle
\section{Introduction}\label{sec:intro}
\noindent Topological materials have been under the spotlight of experimental and theoretical research for years by now, due to their relevance in terms of fundamental physics and their broad spectrum of applications, from  spintronics to quantum computing[\onlinecite{kane-review,zhang-review,ando-review}]. One of the most remarkable features of a topological  phase  is   that edge states localize at the interface with a topologically trivial phase. Indeed several theoretical analysis have shown that such interface states  emerge at the boundaries of  topological insulators (TIs), like the   one-dimensional Su-Schrieffer-Heeger model for polyacetylene [\onlinecite{SSH_PRL,SSH_PRB,kane-lubensky_NaturePhys_2014}]  or the two-dimensional quantum spin Hall systems~[\onlinecite{kane-mele2005a,kane-mele2005b,bernevig_science_2006,zhang_2008,molenkamp-zhang}]. Similarly, as first predicted by Kitaev[\onlinecite{kitaev}],
 at the edges of   topological superconductors[\onlinecite{alicea_review,fujimoto,aguado_review}], realized     in proximized   nanowires (NWs) with Rashba spin-orbit coupling (RSOC)[\onlinecite{vonoppen_2010},\onlinecite{dassarma_2010}], in   ferromagnetic atomic chains deposited   on a superconductor[\onlinecite{yazdani_PRB}], or in two-dimensional TIs proximized by   superconductors and magnets[\onlinecite{fu-kane_PRL_2008,beenakker-akhmerov_2008,crepin-trauzettel-dolcini_PRB_2014}], Majorana quasi-particles (MQPs) appear. These exotic quasi-particles, which are  equal to their anti-particles, are currently considered a promising platform for quantum computing in view of their non-trivial braiding properties and their robustness to charge decoherence effects[\onlinecite{freedman_RMP_2008,alicea_PRX_2016,dassarma-stanescu,akhmerov_2018,kouwenhoven-review}].
\begin{figure}[b]
\centering
\includegraphics[width=8cm,clip]{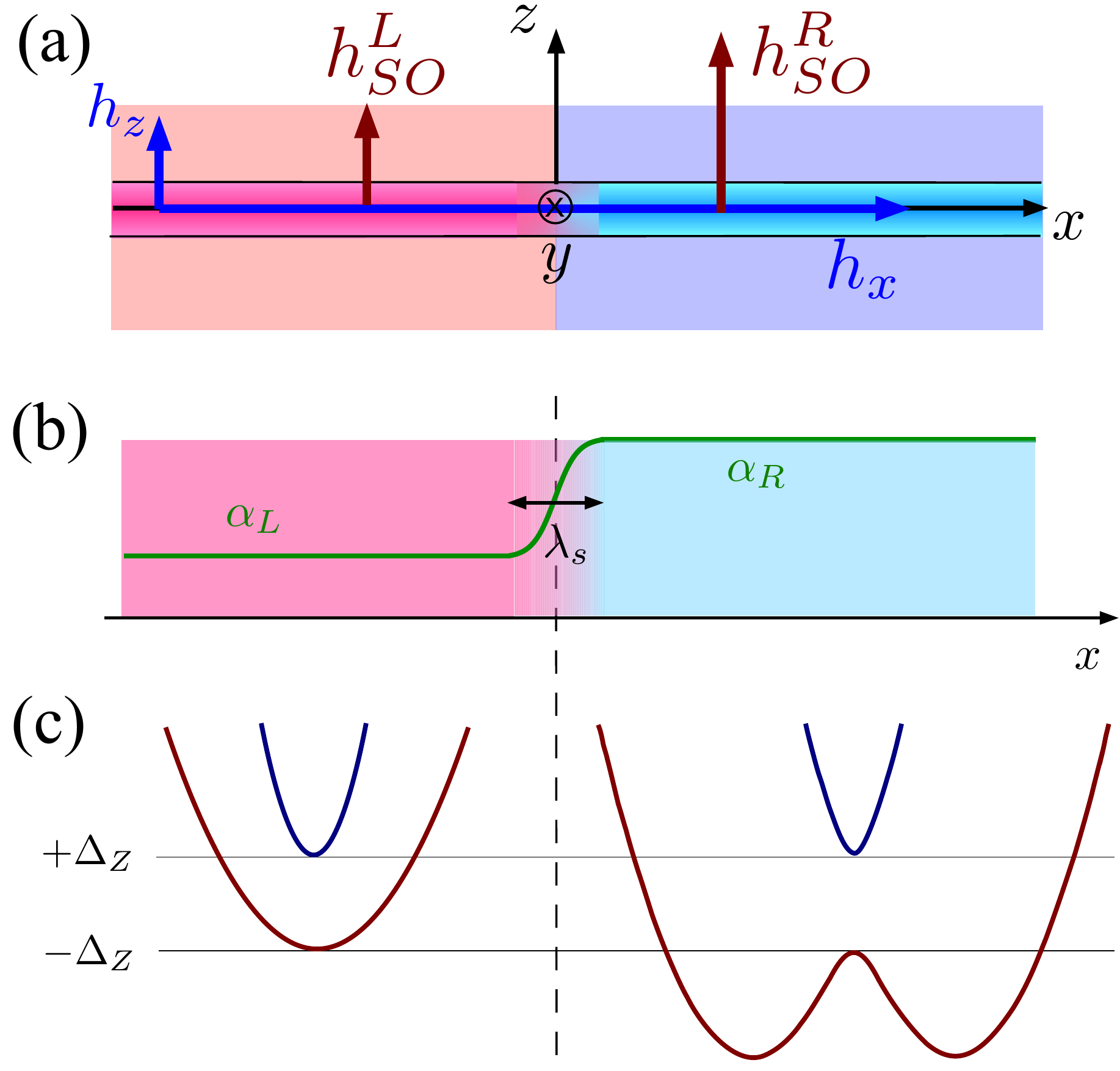}
\caption{(Color online) (a) Top view of a Rashba nanowire deposited on a substrate: the Rashba effective magnetic field $h^{SO}$ is directed along $z$, whereas an actual magnetic field,  externally applied  in the substrate plane, has  components in the $x$-$z$ substrate plane. The NW contains an interface between two regions with different RSOC values. 
(b)  The spatial profile of the RSOC across the interface of the NW, ranging from the bulk values  $\alpha_L$ to $\alpha_R$ over a smoothening lengthscale $\lambda_s$.
(c) Examples of electronic bands related to the bulks of the two interface sides, the left-hand side in the Zeeman dominated regime, and the right-hand side in the Rashba-dominated regime.}
\label{Fig1}
\end{figure}

While in theoretical models a  topological phase is characterized by a well specified range of parameters in the Hamiltonian, when it comes to finding an experimental evidence of such phase  in   a given material, the difficult question is ``how to  distinguish  signatures of a  topological    from a trivial bound state?" As a general criterion,  a topological bound state is stable to perturbations that do not close the gap of the topological phase, while a trivial bound state is not. However, because  in  a given experimental setup the actual parameter range characterizing the topological phase is not known a priori and/or may be relatively narrow, the search for such stable signatures is in general   not a trivial task. For instance, although it is by now commonly accepted that MQPs exist   in RSOC nanowires[\onlinecite{kouwenhoven_2012,liu_2012,heiblum_2012,xu_2012,defranceschi_2014,marcus_2016,marcus_science_2016,kouwenhoven_2018}],    the early observations of a zero-bias conductance peak  stable   to magnetic field and Fermi energy variations were cautiously claimed to be   {\it compatible} with the existence of MQPs. The remark  that such scenario may also be  caused by Kondo effect[\onlinecite{lee-defranceschi_PRL_2012}], disorder[\onlinecite{lee-law_PRL_2012},\onlinecite{dassarma_PRR_2020}] or inhomogeneities[\onlinecite{dassarma_PRB_2017}]  has recently spurred further investigations, which  pointed out that in the topological phase also trivial bound states may be present[\onlinecite{dassarma_PRB_2017,loss_EPJB_2015,trauzettel_2018,frolov_2019,ricco_PRB_2019,tewari_2019,aguado-loss_2019}]. Furthermore, a quite recent analysis[\onlinecite{ronetti-loss_2019}], carried out on a nanowire with homogeneous RSOC and with inhomogeneous magnetic field, showed that at the interface between two magnetic domains with opposite magnetization directions,  bound states appear that are unrelated to the Jackiw-Rebbi  topological states.\\

A more clear evidence of topological bound states requires a spatially resolved analysis. This was done, for instance,  in ferromagnetic atomic chains deposited on a superconductor[\onlinecite{yazdani-science}], where the combined use  of spatially resolved spectroscopic and spin-polarized measurements showed that zero-bias conductance peaks are due to states   localized at the ends of the chain. Yet, the smoking gun enabling one to identify  such states with MQPs is their disappearance in the normal state,   when superconductivity is suppressed. As far as NWs are concerned, it has been pointed out that   MQPs in the topological phase  exhibit an {\it orthogonal} spin density, i.e., a component perpendicular to both the magnetic and spin-orbit fields, localized at the NW ends~[\onlinecite{simon_PRL_2012,black-schaffer_2015,domanski_scirep_2017}].
In order to identify a topological phase in a given system,  it is thus particularly important to understand whether and  when  the {\it topologically trivial} phase may exhibit observables that are spatially localized at the interfaces and that may  mistakenly be  interpreted as a topological signature. So far, this aspect has been analyzed far less than the    topological bound states. 

This paper is meant to bridge this gap. Specifically, we consider the case of a RSOC  NW exposed to a uniform magnetic field, and we analyze the spatial profile of charge and spin densities at the interface between two regions with different values of RSOC, as sketched in Fig.\ref{Fig1}(a). Such type of interfaces emerge quite naturally in any realistic setup, since metallic electrodes or gates are typically deposited on top of a portion of the NW, thereby altering the underneath structure inversion asymmetry characterizing the very RSOC. Furthermore,  the recent advances in various gating techniques, including   gate-all-around approaches, allow   a large tunability of the RSOC constant, possibly even changing the RSOC sign~[\onlinecite{gao_2012,slomski_NJP_2013,wimmer_2015,nygaard_2016,sasaki_2017,tokatly_PRB_2017,loss_2018,tsai_2018}].

Importantly,  on both sides of the interface, the NW that we consider is in the {\it topologically trivial phase},  since no superconducting coupling is included. Furthermore, as the  gap  depends only on the strength of the magnetic field,  it never closes at the interface, since the magnetic field is assumed to be {\it uniform}. Thus, under these conditions  the existence of bound states of    topological origin is ruled out a priori. \\

Our analysis unveils various noteworthy aspects. In the first instance, a bound state  may  appear at the interface. Importantly such bound state, while being not topological, is {\it not} a customary  interface state merely arising from the inhomogeneity of the RSOC. Indeed it can only exist if an external magnetic field is applied orthogonally to the RSOC field direction, and if its intensity fulfills specific conditions with respect to the  two spin-orbit energies characterizing the two NW regions. The conditions of existence and the robustness of the bound state are analyzed in details in terms of different values of RSOC across the interface, including the smoothening length characterizing the crossover between these two values and the presence of a magnetic field component parallel to the RSOC field direction.

Second, we find for realistic values of  chemical potential and temperature that the  orthogonal   spin density exhibits a peak pinned at the interface. Despite the NW is in the topologically trivial phase,  such a peak is relatively robust to other parameter variations. In fact, we show that it  persists even when  the bound state is absent, indicating that in such a case also the continuum states locally modify their spin-texture  to maintain such effect.  

Furthermore, by considering the case of two interfaces, we show that the peaks of the orthogonal   spin density  are opposite at the two ends of the inner NW region, similarly to what occurs for MQPs in the topological phase.  

These results imply that a localized orthogonal spin-density can neither be taken as  a unique signature of a MQP, nor of a topologically trivial bound state. 
However, we argue that it can represent a useful way to indirectly detect spin current differences. Indeed, while the detection of a bulk equilibrium spin current,  which emerges in a homogeneous NW from the correlations between spin and velocity induced by the magnetic and spin-orbit fields[\onlinecite{dolcini-rossi_PRB_2018}], has been elusive so far, any  {\it variation} of equilibrium spin current occurring at the  interface  is precisely related to the orthogonal spin-density peak predicted here.  
 
The paper is organized as follows. In Sec.~\ref{sec-2} we introduce the model and describe the involved energy scales. In Sec.~\ref{sec-3} we present the results concerning the bound state, discussing first the case of a sharp RSOC interface profile in the presence of a magnetic field applied along the NW axis. Then we analyze the more realistic case of a finite smoothening length in the profile, and address  the effect of a magnetic field component parallel to the spin-orbit field direction. In Sec.~\ref{sec-4} we investigate the spatial profile of the charge and spin densities, and analyze specifically the bound state contribution to them. In Sec.~\ref{sec-5} we discuss the interpretation of  our main results, we include the case of two interfaces  and we propose some possible experimental realizations. Finally, in Sec.~\ref{sec-6}  we draw our conclusions.

\section{The model for a SOC interface}
\label{sec-2} 
\subsection{Nanowire Hamiltonian}
Let $x$ denote the longitudinal axis of a NW deposited on a substrate. The NW is characterized by a RSOC, which is assumed to take two different values $\alpha_L$ and $\alpha_R$ on the left and on the  right  side of an interface, respectively [see Figs. \ref{Fig1}(a) and \ref{Fig1}(b)]. This inhomogeneity in the RSOC profile $\alpha(x)$ may result e.g. from the presence of a gate covering only one portion of the NW, or by two different gate voltage values applied to   top/bottom gates or to the substrate. The crossover between $\alpha_L$ and $\alpha_R$ occurs over a smoothening length $\lambda_s$. 
Denoting  by $z$ the direction  of the spin-orbit field $\mathbf{h}^{SO}$, i.e., the effective ``magnetic" field generated by the RSOC [see Fig.\ref{Fig1}(a)], the NW Hamiltonian is 
\begin{equation}\label{H}
\hat{\mathcal{H}}  = \int \hat{\Psi}^\dagger(x)\,H(x)\,  \hat{\Psi}(x)\,dx\quad,
\end{equation}
where
\begin{equation}\label{H-realspace}
H(x)=  \frac{p_x^2}{2 m^*} \sigma_0 -\frac{\left\{ \alpha(x) , p_x\right\}}{2\hbar}  \sigma_z\, \, - \mathbf{h}\cdot \boldsymbol{\sigma} \quad.
\end{equation}
Here   $\hat{\Psi}(x)= ( \hat{\Psi}_\uparrow(x) \,,\,\hat{\Psi}_\downarrow(x)  )^T$ is the electron spinor field, with $\uparrow,\downarrow$ corresponding to spin projections along~$z$, $p_x=-i\hbar \partial_x$ is the momentum operator, $m^*$ the NW effective mass, 
$\sigma_0$ the $2 \times 2$ identity matrix, and $\boldsymbol{\sigma}=(\sigma_x,\sigma_y,\sigma_z)$ are the Pauli matrices. For definiteness, we take  the location of the interface at $x=0$.
The anticommutator in Eq.(\ref{H-realspace}) is necessary since  $p_x$ does not commute with the inhomogeneous RSOC $\alpha(x)$
~[\onlinecite{sanchez_2006},\onlinecite{sanchez_2008}]. The last term in Eq.(\ref{H-realspace}), where $\mathbf{h}=g \mu_B \mathbf{B}/2$, describes the Zeeman coupling with  an  external uniform magnetic field $\mathbf{B}=(B_x,0,B_z)$ applied   in the substrate plane, with $\mu_B$ denoting the Bohr magneton and $g$ the NW Land\'e  factor. 
It is useful to decompose the magnetic gap energy vector as $\mathbf{h}={h_x} \mathbf{i}_x+{h_z} \mathbf{i}_z$, where $h_x$ and $h_z$ denote the components parallel and perpendicular to the nanowire axis $x$, i.e., perpendicular and parallel to the Rashba spin-orbit field direction $z$, respectively  [see Fig.\ref{Fig1}(a)]. Although for most of our analysis we shall focus on the case of the magnetic field   directed along the nanowire axis $x$, we shall also discuss the effects of the component  $h_z$ parallel to $\mathbf{h}^{SO}$.  

\subsection{Energy scales}
In order to describe the results about the inhomogeneous RSOC profile at the interface, it is first worth pointing out the energy scales involved in the problem.
\subsubsection{The homogeneous NW}
Let us start by briefly summarizing the case of a homogeneous profile $\alpha(x) \equiv \alpha$ in Eq.(\ref{H-realspace}),  for an infinitely long NW. In such case the Hamiltonian (\ref{H-realspace})   commutes with~$p_x$, and the spectrum reads~[\onlinecite{dolcini-rossi_PRB_2018,sanchez_2006,sanchez_2008}]
\begin{equation}\label{spectrum}
E_\pm(k)=  \, \varepsilon^0_k\pm \sqrt{h_x^2+(\alpha k +{h_z})^2}\quad,
\end{equation}   
where $\varepsilon^0_k=\hbar^2 k^2/2 m^*$ is the customary parabolic spectrum in the absence of RSOC and magnetic field. 
The spectrum (\ref{spectrum}) describes two bands    separated by a minimal gap $2\Delta_Z$, where the quantity
\begin{equation}
\Delta_Z=|h_x| \label{EZ-def}
\end{equation}
shall be henceforth called the magnetic gap energy.  Moreover, the RSOC $\alpha$ identifies  the spin-orbit wavevector
$k_{SO}= m^* |\alpha|/\hbar^2$, 
which characterizes, in the absence of external magnetic field,  the two degenerate  minima $E(\pm k_{SO})=-E_{SO}$ of the spectrum, where
\begin{equation}\label{ESO-def}
E_{SO}=\frac{m^*\alpha^2}{2\hbar^2}  = \frac{\hbar^2 k_{SO}^2}{2 m^*} 
\end{equation}
is called the spin-orbit energy. 

In the case  $h_z=0$ the magnetic field is directed along~$x$, i.e., orthogonal to the RSOC field, the spectrum (\ref{spectrum}) is symmetric $E_\pm(-k)=E_\pm(+k)$. Two regimes can be identified: 
(a) in the {\it  Zeeman-dominated regime} ($\Delta_Z > 2E_{SO}$)  both bands have a minimum at $k=0$, which takes values $E_{\pm}^{\rm min}=\pm \Delta_Z$, respectively.\\
(b) in the {\it  Rashba-dominated regime} ($\Delta_Z < 2E_{SO}$),  the upper band still has a minimum $E_{+}^{\rm min}=+\Delta_Z$ at $k=0$, while the lower band acquires two lower and degenerate minima $E_{-}^{\rm min}=-E_{SO}-\Delta_Z^2/4 E_{SO}$ occurring at $k=\pm k^{min}$, with 
\begin{equation}
k^{min}= k_{SO} \sqrt{1-\Delta_Z^2/4 E_{SO}^2} \quad.
\end{equation}

When a component $h_z \neq 0$ parallel to the RSOC field is also present, the minimal gap $2\Delta_Z$ between the two bands occurs at $k=-h_z/\alpha$  and the spectrum is no longer symmetric $E_\pm(-k) \neq E_\pm(+k)$.\\

The eigenfunctions related to the spectrum (\ref{spectrum}) read
\begin{equation}
\label{eigen-homo}
\psi_{k\pm}(x)=w_{k \pm} \exp[i k x]/\sqrt{\Omega} \quad,
\end{equation} 
with $\Omega$ denoting the system length. They describe plane waves with spinors
\begin{eqnarray}
\label{eigenvectors}
w_{k-}= \left(
\begin{array}{c}
\cos \frac{\theta_k}{2} \\  \\ 
 \sin \frac{\theta_k}{2}\,  
\end{array}\right)\hspace{0.5cm} 
w_{k +}= \left(
\begin{array}{c}
-  \sin \frac{\theta_k}{2}\\  \\ 
\cos \frac{\theta_k}{2}
\end{array}\right)\,, \hspace{0.5cm} 
\end{eqnarray}
whose spin orientation $\mathbf{n}(k) \equiv \left( \sin\theta_k \,,0 \,,\, \cos\theta_k \right)$ lies  on the $x$-$z$ substrate plane and forms with the $z$-axis an angle~$\theta_k \in [-\pi , \pi ]$. The latter, defined through
\begin{equation}\label{thetak-def}
\left\{\begin{array}{lcl}
\cos \theta_k  &=&\displaystyle  \frac{\alpha k+{h_z}}{\sqrt{(\alpha k +{h_z})^2+h_x^2}}\\  
\sin \theta_k  &=&\displaystyle \frac{{h_x}}{\sqrt{(\alpha k +{h_z})^2+h_x^2}}
\end{array}\right.   \quad,
\end{equation}
depends on the wavevector $k$,  the magnetic field  and   the RSOC $\alpha$. 
In particular, it is worth recalling that  in the case of a magnetic field along the NW axis ($h_z=0$) and in the deep Rashba-dominated regime ($\Delta_Z \ll 2 E_{SO}$)  the states with energy inside the magnetic gap mimic the helical edge states of the quantum spin Hall  effect. Indeed their spin orientation, determined mainly by the RSOC, is opposite for right- and left-moving electrons, whose helicity is determined by the {\it sign} of the RSOC~$\alpha$. This is precisely the most suitable regime for the topological phase to be induced by an additional $s$-wave superconducting coupling~[\onlinecite{vonoppen_2010},\onlinecite{dassarma_2010},\onlinecite{lutchyn_2012},\onlinecite{loss_PRB_2017}].

\subsubsection{The NW with a RSOC interface}
\label{sec-2-B-2}
When an interface separates two portions of a NW characterized by  two different values $\alpha_L$ and $\alpha_R$  of RSOC  [see Fig.\ref{Fig1}(b)], the momentum $p_x$ does not commute with the Hamiltonian  characterized by an  inhomogeneous $\alpha(x)$-profile, and the spectrum cannot be labeled by a wavevector $k$. Before attacking the inhomogeneous problem  in the next section, it is worth  identifying the  energy scales and the  possible scenarios 
one can  expect  in the interface problem  from a preliminary analysis of the    bulks of the two regions across the RSOC interface.  To begin with, the two bulk values  $\alpha_L$ and $\alpha_R$ of the two NW regions lead to  two spin-orbit energies (\ref{ESO-def})
\begin{equation} \label{ESO-nu-def}
E_{SO,\nu}=\frac{m^*\alpha^2_\nu}{2\hbar^2} \hspace{2cm} \nu=R/L \quad.
\end{equation}
Without loss of generality, we shall choose the  RSOC with higher magnitude $|\alpha|$ on the right-hand side, and  we can set it to a positive value, $\alpha_R>0$, whereas the RSOC on the left-hand side is allowed to take any value in the range $-\alpha_R \le \alpha_L \le \alpha_R$~ [\onlinecite{nota-su-parity}]. Correspondingly, one has $E_{SO,L} \le E_{SO,R}$. The fact that the magnetic field is uniform has important consequences, which are easily illustrated in the case $h_z=0$: First,  in the bulk of each region  the gap between the bands is always given by $2\Delta_Z$, regardless of   the regime (Rashba- or Zeeman-dominated) of  each interface side. Secondly,  the overall minimum  of the two energy band bottoms  is determined by the band bottom of the side with higher spin-orbit energy, i.e., the right-hand side, and is thus given by
\begin{equation} 
E^{\rm min}_{band}=\left\{ 
\begin{array}{ll} 
-\Delta_Z &   \mbox{if } \Delta_Z>2 E_{SO,R}  \\   & \\
 -E_{SO,R}\left(1+\frac{\Delta_Z^2}{4 E^2_{SO,R}} \right)&   \mbox{if } \Delta_Z<2 E_{SO,R} 
\end{array}\right. \label{Emin-band}
\end{equation}
With these notations, if the right-side is in the Zeeman-dominated regime,  so is the left-hand side, whereas if the right-side is in the Rashba-dominated regime the left-hand side can be either in the Rashba- or in the Zeeman-dominated regime. There can thus be only three possible regime combinations: (i) $E_{SO,L} \le E_{SO,R} \le \Delta_Z/2$, where both sides are Zeeman-dominated; (ii) $\Delta_Z/2 \le E_{SO,L} \le E_{SO,R}$, where both sides are Rashba-dominated; (iii) $E_{SO,L} \le \Delta_Z/2  \le E_{SO,R}$, where the left-side is Zeeman-dominated while the right-side is Rashba-dominated. The bands of the latter case are illustrated as an example in Fig.\ref{Fig1}(c).\\

\section{Bound state and its stability}
\label{sec-3}
In this section we focus on the inhomogeneous interface problem. By diagonalizing the  inhomogeneous Hamiltonian, with methods to be described here below, we find that its spectrum always exhibits a continuum branch, whose bottom $E^{\rm min}_{cont}$ coincides  with the minimal band energy   obtained in Eq.(\ref{Emin-band})  from the comparison of bare bulk spectra. However, for some parameter range  (see below), the spectrum also displays an additional eigenvalue~$E_{bs}$, lying 
 {\it below} the continuum spectrum $E^{\rm min}_{cont}$. The related eigenfunction   exhibits an evanescent behavior for $|x|\rightarrow \infty$. When such bound state   exists, we define its positive `binding energy' as
\begin{equation} \label{Eb-def}
E_b=E^{\rm min}_{cont}-E_{bs}\, >0\quad.
\end{equation}
Here below we now analyze the conditions for its existence.

\subsection{The case of a sharp interface}
Let us start by analyzing the existence of the bound state  in the case of a sharp interface, where  the smoothening length $\lambda_s \rightarrow 0$ vanishes and the profile can be assumed as 
\begin{equation}\label{alpha-sharp}
\alpha(x)=\alpha_L\theta(-x)+\alpha_R\, \theta(x)
\end{equation}
with $\theta$ denoting the Heaviside function. In this case the eigenfunctions of the inhomogeneous problem can be obtained analytically by combining the eigenstates (\ref{eigen-homo}) of the homogeneous problem in each side and by matching them appropriately at the interface. In particular, since  bound states are eigenstates with evanescent wavefunction for $|x|\rightarrow \infty$, they are obtained requiring that the wavevector $k$ acquires an imaginary part. Details of such calculation can be found in Appendix~\ref{AppA}.  

By keeping one side of the junction as a reference, e.g. the right-hand side  where the bulk spin-orbit energy   is maximal, the problem  can be formulated   in terms of dimensionless parameters, namely  the RSOC   ratio $\alpha_L/\alpha_R \in [-1\,, 1]$ and the energy ratios $E_b/E_{SO,R}$ and $\mathbf{h}/E_{SO,R}$ to the maximal spin-orbit energy $E_{SO,R}$. We shall focus here below on the case where the applied magnetic field is directed only along the nanowire axis~$x$, $\mathbf{h}=h_x \mathbf{i}_x$, i.e., orthogonally to the Rashba spin-orbit field, while the effects of a parallel magnetic field component $h_z$ will be discussed later.

\begin{figure}[t]
\centering
\includegraphics[width=8cm]{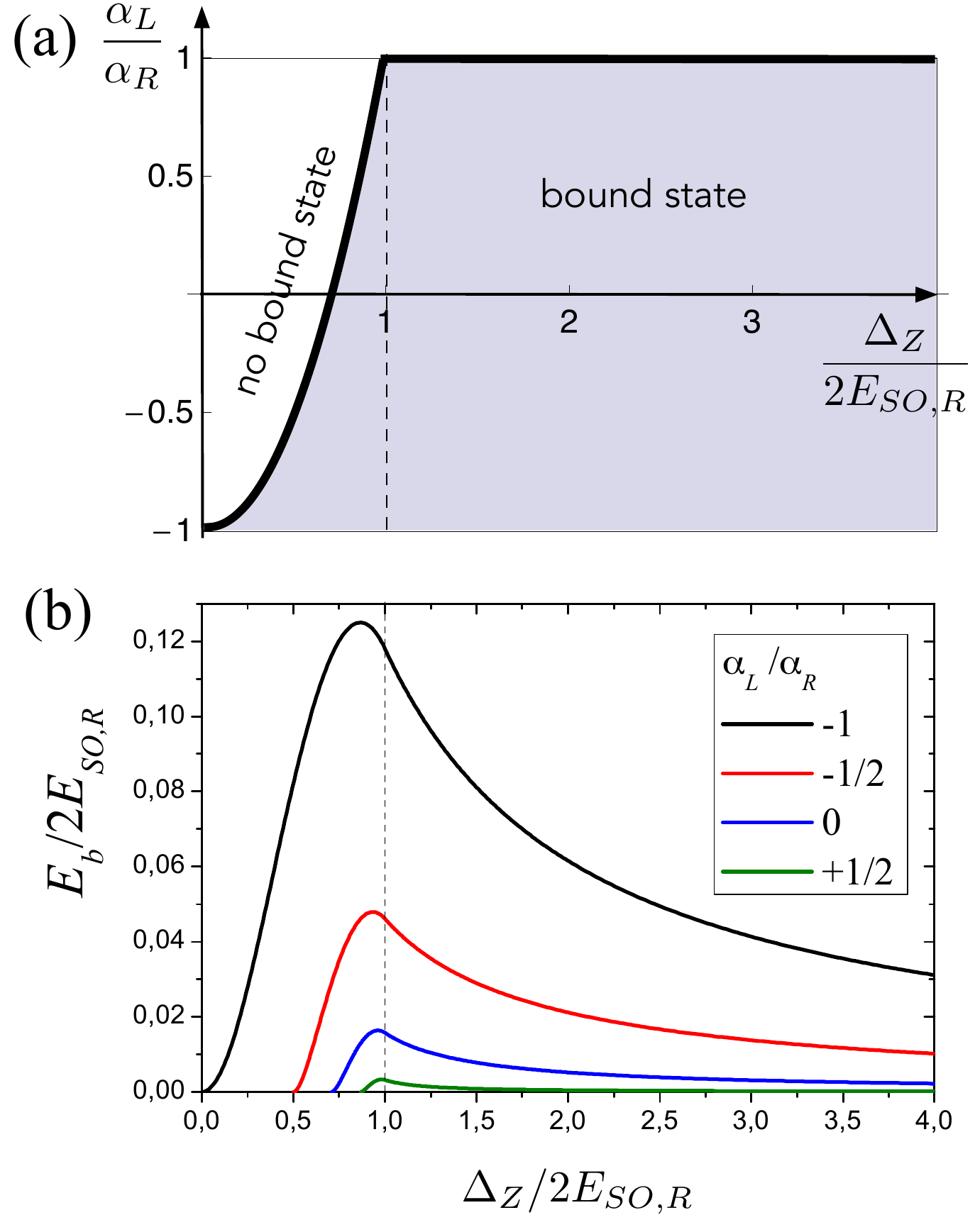}
\caption{(Color online)  The case of a sharp profile interface Eq.(\ref{alpha-sharp}). (a) The phase diagram for the existence of the bound state is shown  as a function of the  magnetic gap energy (in units of twice the maximal spin-orbit energy $2E_{SO,R}$) and of the ratio between the two RSOC values across the interface. The thick black line identifies the transition curve, where the binding energy vanishes. The vertical thin dashed line indicates the crossover value from the Rashba-dominated to the Zeeman-dominated regime for the right-side of the interface. (b) The binding energy $E_b$ of the bound state as a function of $\Delta_Z/2 E_{SO,R}$ for four different values of the RSOC ratio across the interface.}
\label{Fig2}
\end{figure}

The results are presented in Fig.\ref{Fig2}. In particular, Fig.\ref{Fig2}(a) displays the phase diagram of the existence of the bound state.   For a sufficiently strong   magnetic field, $\Delta_Z > 2E_{SO,R}$, i.e., when both NW sides are in the Zeeman-dominated regime, the bound state always exists, while for $\Delta_Z < 2E_{SO,R}$, where the NW right side  is in the Rashba-dominated regime, the bound state may or may not exist. 
In particular, for $\Delta_Z=0$ (no external magnetic field), the bound state never exists, regardless of the ratio of the two RSOC values across the interface. This shows that the bound state, although it has no topological origin, it is   not an intrinsic interface state like the ones occurring at a customary semiconductor interface. 
The thick black in Fig.\ref{Fig3}(a) denotes the transition curve for the existence of the bound state, and corresponds to the vanishing of the binding energy, $E_b=0$. In particular,  the parabolic curve for $\Delta_Z/ 2E_{SO,R}<1$ is described by the equation
\begin{equation}
\frac{\Delta_Z^\star}{2E_{SO,R}}=\sqrt{\frac{1+\alpha_L/\alpha_R}{2}} \label{EZ-star} \quad,
\end{equation}
while the upper horizontal line  corresponds to the homogeneous NW in the Zeeman-dominated regime, where the bound state does not exist, as is obvious to expect.
\noindent Then, Fig.\ref{Fig2}(b) shows, for  four different values of the ratio $\alpha_L/\alpha_R$, the  behavior  of the binding energy $E_b$  as a function of the ratio $\Delta_Z/ 2E_{SO,R}$. Several features are noteworthy. 

First, in all cases the binding energy exhibits a non-monotonic behavior as a function of the magnetic gap energy, with a maximum $E_b^{\rm max}$ occurring for a magnetic gap energy slightly below the transition value $\Delta_Z=2 E_{SO,R}$  between the Rashba- and Zeeman-dominated regime of the right-hand side, highlighted by the vertical dashed line as a guide to the eye. 

Secondly, the bound state energy strongly depends on the ratio $\alpha_L/\alpha_R$ of the two  RSOC values, and is typically much higher when the RSOC changes sign across the interface. In particular, the
optimal condition for the existence of the bound state is   $\alpha_L/\alpha_R=-1$, i.e., when the RSOC takes  {\it equal and opposite} values of two sides: In this situation not only the bound state always exists, its  binding energy is also higher than any other case. For these reasons, we shall henceforth term such case the `optimal configuration'. In particular, it can be shown that, for weak applied field ($\Delta_Z \ll 2 E_{SO,R}$) the binding energy of the optimal configuration behaves as 
$E_b  \simeq  \Delta_Z^2 / 4E_{SO,R}$ while for strong field ($\Delta_Z \gg 2 E_{SO,R}$) one finds $E_b \simeq   E^2_{SO,R}/ 2 \Delta_Z$.

Third, for all other cases ($-1< \alpha_L/\alpha_R<1$)  the bound state exists only if the magnetic gap energy overcomes a minimal threshold value, which precisely corresponds to the transition curve of  Fig.\ref{Fig2}(a) described by Eq.(\ref{EZ-star}). The threshold of the magnetic gap energy increases as the RSOC ratio $\alpha_L/\alpha_R$ increases from the negative value~$-1$ to the value $+1$, corresponding to the homogeneous case.  Furthermore, the following `rule of thumb' can be inferred: when the  band bottoms of the two interface sides are equal, the bound state certainly exists. Indeed a close inspection of Fig.\ref{Fig2} shows that this certainly occurs in these two situations: (i) when  $\Delta_Z/2 E_{SO,R} >1$, i.e., when both   sides are in the Zeeman-dominated regime and their band bottoms are both equal to $-\Delta_Z$; (ii) when $\alpha_L=-\alpha_R$, i.e., when the two spin-orbit energies (\ref{ESO-nu-def}) are equal,   both sides are   in the same regime (Rashba- or Zeeman-dominated) and thus have the same band bottoms. In all other cases the existence of the bound state depends on the specific energy ratios.

Finally, even when the bound state exists, its binding energy can be quite small. For instance, the maximal binding energy in the case where $\alpha_L/\alpha_R=1/2$    is about 25 times smaller  than the maximal value in the optimal case $\alpha_L/\alpha_R=-1$. Similarly, even in the regime $\Delta_Z/2 E_{SO,R} >1$   the binding energy decreases with increasing magnetic field.

\subsection{Effects of smoothening length}
\begin{figure}[t]
\centering
\includegraphics[width=8cm]{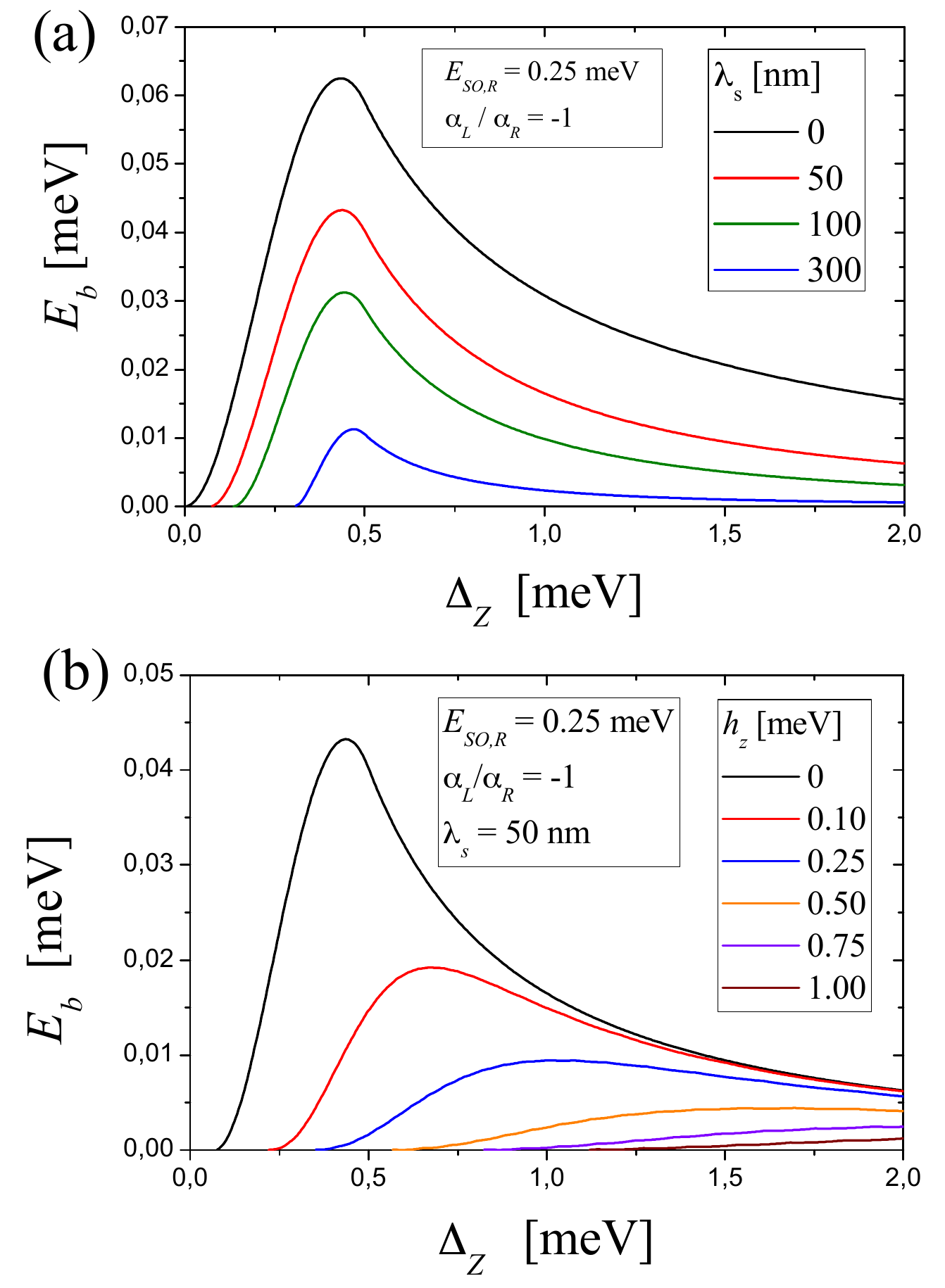}\\
\caption{(Color online) The binding energy as a function of the magnetic gap energy, for an interface with $\alpha_L=-\alpha_R$, with $E_{SO,R}=E_{SO,L}=0.25\,{\rm meV}$. (a) The effects of a smoothening length. (b) Effects of a magnetic field component $h_z$ parallel to the spin-orbit field on the binding energy, for a fixed smoothening length $\lambda_s=50\,{\rm nm}$.}
\label{Fig3}
\end{figure}

In any realistic system the crossover between  two RSOC bulk values occurs over a finite smoothening length $\lambda_{s}$. To include such effect we now assume   the  following profile function  
\begin{equation}\label{alpha(x)}
\alpha(x)=\frac{\alpha_R+\alpha_L}{2}+\frac{\alpha_R-\alpha_L}{2}\, \mathrm{Erf}\left(\frac{\sqrt{8} \,x }{ \lambda_{s}}\right) \, ,
\end{equation}
which varies from $\alpha_L$ to $\alpha_R$ up to 2\% within the lengthscale  $\lambda_s$. In Eq.(\ref{alpha(x)}) Erf denotes the error function. Although in the presence of such smoothened profile  the model cannot be solved analytically, it can be approached by an exact numerical diagonalization of  the Hamiltonian (\ref{H-realspace}), with a   method similar  to the one introduced in Ref.[\onlinecite{dolcini-rossi_PRB_2018}], whose details specific to the   profile (\ref{alpha(x)}) are summarized in App.\ref{AppB}. 
Instead of expressing the results in terms of dimensionless parameters, we now choose to fix the parameters to realistic setup values. For definiteness, we consider  the case of a ${\rm InSb}$ NW, with an effective mass $m^*=0.015\,m_e$ and a maximal spin-orbit energy $E_{SO,R}=0.25\,{\rm meV}$.  Furthermore, in order to appreciate the effects of the smoothening length, we focus on the  case of the optimal configuration $\alpha_R/\alpha_L=-1$. The results,  displayed in Fig.\ref{Fig3}(a), show the binding energy  as a function of the magnetic gap energy $\Delta_Z$ for four different values of the smoothening length. As one can see, while for the ideal case $\lambda_s \rightarrow 0$ (sharp profile) the bound state always exists, for any finite smoothening length the bound state only appears above a threshold value for the Zeeman field.  
For sufficiently strong applied magnetic field (Zeeman-dominated regime) the bound state always exists. However, the binding energy exhibits an overall suppression for increasing $\lambda_s$.
 These effects can be understood be realizing that a crossover from  $-\alpha_R$ to $\alpha_R$ in the RSOC profile occurring over a finite smoothening length can, to a first approximation, be considered as a stair-like sequence of smaller sharp $\alpha$-steps. As the  analysis carried out above on the sharp profile  indicates (see Fig.\ref{Fig2}), in the case of a  non-optimal jump $\alpha_L > -\alpha_R$, a threshold value for $\Delta_Z$ does exist   and the binding energy is reduced.   In summary,   a finite smoothening length $\lambda_s$   broadens  the white portion of the sharp-profile phase diagram Fig.\ref{Fig2}(a) where the bound state does not exist, and suppresses the binding energy.

\subsection{Effects of a parallel field component}
So far, we have analyzed   cases where the magnetic field~$h_x$ is directed along the NW. Here we want to discuss the effect of a magnetic field component $h_z$ parallel to the spin-orbit field. We first point out that, for   $h_z \neq 0$ and $h_x=0$, i.e., for a magnetic field  directed  purely along the spin-orbit field direction $z$, the eigenvalue problem for the Hamiltonian~(\ref{H-realspace}) completely decouples in the two spin-$\uparrow$ and spin-$\downarrow$ components, and it can be shown  that  the bound state does not exist (see App.\ref{AppA}). The orthogonal field component $h_x$ is thus a necessary, though not sufficient, condition for the bound state to exist. One can then analyze how the parallel field component $h_z$ modifies the existence of the bound state, for a fixed value of $h_x \neq 0$. To this purpose, we focus again on a {\rm InSb} NW, with an  optimal configuration
 $\alpha_R=-\alpha_L>0$, and we take a  realistic smoothening length $\lambda_s=50\,{\rm nm}$. The result, displayed in Fig.\ref{Fig3}(b), shows that the  presence of   an additional parallel field component $h_z$ modifies the dependence of the binding energy $E_b$ as a function of the magnetic gap energy $\Delta_Z$, especially by increasing the threshold value $\Delta_Z^\star$ at which the bound state starts to exist. Similarly to the case of the smoothening length, the binding energy values are quite reduced as compared to the case $h_z=0$.

\section{Charge and spin density spatial  profiles}
\label{sec-4}
In the previous section we have discussed the existence and the robustness of the bound state, which is a spectral feature. Here we wish to analyze spatial behavior of   physical observables, namely the charge and  spin densities, described by the operators
\begin{eqnarray}
\hat{n}(x)&=&  {\rm e}\,\hat{\Psi}^\dagger(x) \,   \hat{\Psi}^{}(x)  \label{charge-density-def} \\
\hat{\mathbf{S}}(x) &=&\frac{\hbar}{2} \,\hat{\Psi}^\dagger(x) \, \boldsymbol\sigma\, \hat{\Psi}^{}(x)\quad, \label{spin-density-def} 
\end{eqnarray}
respectively, where ${\rm e}$ denotes the electron charge. The presence of the interface makes the NW an inhomogeneous system, and we aim to investigate the spatial profile of the equilibrium expectation values 
\begin{eqnarray}
\rho(x) &\equiv &\frac{1}{\rm e} \langle \hat{n}(x) \rangle_\circ \label{rho-def} \\
\mathbf{s}(x) &\equiv & \frac{2}{\hbar}\langle \hat{\mathbf{S}}(x) \rangle_\circ \label{s-def}
\end{eqnarray}
 with  a particular focus on  their behavior    near the interface. Details about the computation of such expectation values can be found in App.\ref{AppB}. Before presenting our results, a few general comments are in order.\\
  
{\it Chemical potential and Temperature.} The equilibrium distribution determining the expectation values (\ref{rho-def}) and (\ref{s-def})  is characterized by a well defined value of chemical potential $\mu$ and temperature $T$. As pointed out above, the whole spectrum of the inhomogeneous  Hamiltonian (\ref{H-realspace}), which we obtain by an exact numerical diagonalization, consists of a continuum spectrum, related to extended propagating states, and possibly (if present) a bound state, energetically lying below the continuum and corresponding to a state localized at the interface.
At equilibrium, and ideally at zero temperature, all states (localized or extended) with energy up to the chemical potential   $\mu$ are filled up, and contribute to determine the equilibrium expectation values $\rho(x)$ and $\mathbf{s}(x)$, while at finite  temperature  the Fermi function is smeared over a range $k_B T$ around the chemical potential. 
We shall choose for $T$ and $\mu$  realistic values of  low-temperature experimental setups involving NWs, namely $T=250\,{\rm mK}$  and
$\mu=0$, corresponding to the energy value in the middle of the magnetic gap  [see Fig.\ref{Fig1}(c)]. This is the situation, for instance, where  the Fermi energy states of a NW in the Rashba-dominated regime mimic the helical states of a quantum spin Hall system.

{\it Orthogonal spin density.} Concerning the spin density~$\mathbf{s}(x)$ in Eq.(\ref{s-def}), we shall specifically focus on $s_y$ component, which we shall refer to as the {\it orthogonal spin density}, since it is orthogonal to   the $x$-$z$ plane identified by the applied magnetic field and the spin-orbit field. The interest in analyzing the profile of $s_y(x)$  stems from a comparison with the topological phase. Indeed it has been predicted [\onlinecite{simon_PRL_2012,black-schaffer_2015,domanski_scirep_2017}] that the MQPs appearing at the   ends of a proximitized NW in the topological phase, are precisely characterized by a non-vanishing expectation value $s_y$.
However, we shall show here below that such orthogonal spin density  already appears in the NW interface problem, where the NW is certainly in the topologically trivial phase, so that it cannot be considered as a signature of a MQP.  \\

{\it Full vs. bound state contribution.} Bound states and orthogonal spin density $s_y$ share two   properties. First, both can only exist at an interface, i.e., in the presence of inhomogeneities. Indeed,  in the bulk of a homogeneous NW, $s_y$ vanishes   since the spin orientation of each electron lies in the $x$-$z$ plane  [see Eqs.(\ref{eigenvectors})-(\ref{thetak-def})].
Second,  just like the bound state, $s_y$ may only exist if both a magnetic field component $h_x$ and the spin-orbit field are present.   Indeed if $h_x=0$ (or $\alpha=0$) the electron spin is directed, along $z$ (or $x$)  for all  states. In view of such common features, one is naively tempted to conclude that an orthogonal spin density is necessarily ascribed to the presence of the bound state. However, this is not the case. To this purpose, we shall illustrate below two types of spatial profiles. First, we shall show  
 the   actual equilibrium values $\rho(x)$ and $s_y(x)$ [see Eqs.(\ref{rho-def}) and (\ref{s-def})], which can be referred to as the `full'  density and orthogonal spin density profiles, as they result  from   contributions of all states, with the customary weight given by the Fermi   function. In particular, since we focus on the low temperature regime, the latter   essentially amounts to the contribution of all states occupied up to the chemical potential $\mu$. Then, we shall also provide the profiles $\rho_{bs}(x)$ and $s_{y,bs}(x)$ describing the contribution to $\rho(x)$ and $s_y(x)$ due to the localized bound state only  [see  App.\ref{AppB} for details]. 
  
This distinction enables us to show that an orthogonal spin density peak, besides being no evidence for a MQP, may also not originate from any bound state.

\begin{figure*} 
\centering
\includegraphics[width=\textwidth]{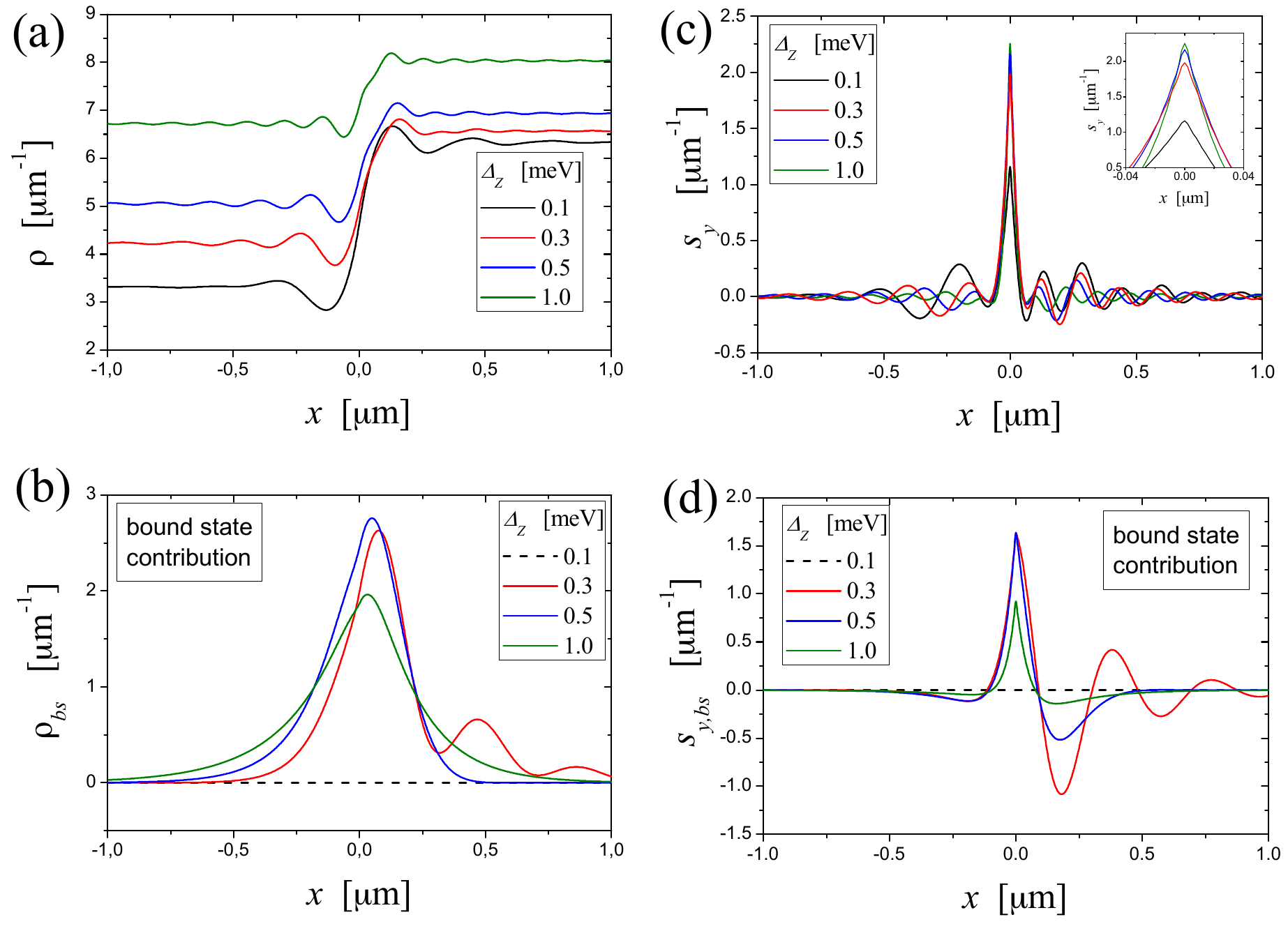}\\
\caption{(Color online) Spatial profiles of charge density and orthogonal spin density for a sharp  interface profile Eq.(\ref{alpha-sharp}) with $\alpha_L/\alpha_R=-1/2$ and $E_{SO,R}=0.25\, {\rm meV}$. The four different curves in each panel refer to  four different values of the magnetic gap energy $\Delta_Z=(0.1, 0.3, 0.5, 1.0)\,{\rm meV}$. (a)  The actual equilibrium density profile $\rho(x)$ [see Eq.(\ref{rho-def})]. (b) The bound state contribution $\rho_{bs}(x)$ to the density $\rho(x)$. For $\Delta_Z=0.1\,{\rm meV}$  the bound state does not exist  and yields a vanishing contribution (black dashed curve). 
 Panel (c) describes the full orthogonal spin density $s_y$ Eq.(\ref{s-def}) (with the inset magnifying the peaks) while panel (d) describes  the related bound state  contribution $s_{y,bs}$.}
\label{Fig4}
\end{figure*}

\subsection{The case of a sharp profile with an orthogonal magnetic field}
Let us start our analysis   from the case of a sharp profile interface and a magnetic field applied along the NW axis. As an illustrative example, we consider an interface with $\alpha_L/\alpha_R=-1/2$, which implies $E_{SO,L}=E_{SO,R}/4$ [see Eq.(\ref{ESO-nu-def})], and we choose a  value  of $E_{SO,R}=0.25\, {\rm meV}$ for the maximal spin-orbit energy.  \\

Figure~\ref{Fig4}(a) shows the  full equilibrium density  Eq.(\ref{rho-def}),   for four different values of the magnetic gap energy $\Delta_Z$ of the applied magnetic field $h_x$. Its spatial profile  $\rho(x)$  exhibits a crossover at the interface  $x=0$ between two different bulk density values. The  density increases towards the  right-hand side, namely the region with higher spin-orbit energy, whose band bottom  is lower than  on the left-hand side with lower spin-orbit energy, as observed    above in Sec.\ref{sec-2-B-2}. This indicates that a higher spin-orbit energy  has a similar effect on the density as a lower gate voltage bias.  

In Fig.\ref{Fig4}(b) we have singled out the contribution   $\rho_{bs}$ due to the bound state only. Differently from $\rho(x)$,
the profile of $\rho_{bs}(x)$  is localized only around the interface and is  dramatically sensitive to the value of $\Delta_Z$. Indeed, as can be deduced  from Eq.(\ref{EZ-star}),    the  minimal threshold for the appearance of the bound state is, for the chosen parameters,  $\Delta_Z^\star= E_{SO,R} =  0.25\, {\rm meV}$. For  values $\Delta_Z>\Delta_Z^\star$ [red, blue and green curves in Fig.\ref{Fig4}(b)], where the bound state exists,  a comparison of the height of the peak of $\rho_{bs}$ with the profile of the full $\rho$  [Fig.\ref{Fig4}(a)] suggests that the    increase of $\rho$ across the interface is mainly due to the presence of the bound state. However, for a  magnetic gap energy $\Delta_Z<\Delta_Z^\star$  [black dashed curve in Fig.\ref{Fig4}(b)] $\rho_{bs}$ is vanishing because the bound state is absent. Note the striking difference from the behavior of the full $\rho(x)$ across the interface [Fig.\ref{Fig4}(a)], which is instead qualitatively very similar for all values of the magnetic gap energy $\Delta_Z$.   In conclusion,  the increase of the profile of $\rho$ at the interface is not necessarily ascribed to a bound state. 
This sounds reasonable, since the electron density is a bulk property receiving contributions from all states up to the chemical potential, and the bound state is just one of such contributions.  The same reasoning holds for the $s_x$ component of the spin density [see Eq.(\ref{s-def})], which is also a bulk quantity, due to the applied magnetic field~$h_x$.

Let us now turn to consider the spin density~$s_y$. Differently from  $\rho$ and from $s_x$,  the  orthogonal spin density $s_y$ is vanishing   in the bulk of a homogeneous NW, as observed above.  Thus, $s_y$  can  only exist (if it does) in  the presence of inhomogeneities, and one could naively expect that it is the hallmark of the presence of a bound state localized at the interface. The profile of the full $s_y$, plotted in Fig.\ref{Fig4}(c), provides two important insights. First, a peak of the orthogonal spin density $s_y$ does exist, even if the NW is in the topologically trivial phase, implying that it cannot be  a unique signature of MQP. Second, the 
central  peak at the interface is weakly sensitive to the values of the magnetic gap energy  $\Delta_Z$. This is in striking contrast to  the behavior of  the bound state contribution $s_{y,bs}$, shown in Fig.\ref{Fig4}(d), which  is again strongly dependent on the magnetic field. In particular, just like   the density $\rho_{bs}$, for  weak Zeeman field  $s_{y, bs}$ vanishes since the bound state is absent (dashed curve), while for higher magnetic field its broadening depends on $\Delta_Z$.
These results show that a localized peak of orthogonal spin density $s_y$ is not necessarily ascribed to the presence of a bound state, neither topological nor trivial. \\

Before concluding this subsection, a few further comments about Fig.\ref{Fig4} are in order. 
We observe that, while the spatial profile of the bound state density $\rho_{bs}$  [panel (b)] is smooth, the profile of  $s_{y,bs}$  [panel (d)] exhibits a cusp at the interface. This difference originates from the boundary conditions induced by the sharp profile (\ref{alpha-sharp}), which cause spin-diagonal observables like $\rho$ and $s_z$ to have continuous derivatives, while spin off-diagonal observables like $s_x$ and $s_y$ to exhibit a cusp at the interface (see App.\ref{AppA}).  Moreover, for $\Delta_Z=0.3\,{\rm meV}$, i.e., slightly above the threshold $\Delta_Z^\star=0.25\,{\rm meV}$, the  profiles of the bound state contributions exhibit a slowly decaying oscillations on the right-hand side, since the bound state wavefunction is characterized by a complex wavevector~$k$ on such side. In contrast, for $\Delta_Z=0.5\,{\rm meV}$ and $\Delta_Z=1.0\,{\rm meV}$ the wavevector is purely imaginary, and the bound state density profile has an exponential decay without oscillations. Finally, the  peak of the orthogonal spin density  $s_{y,bs}$ has a narrower extension than the one of $\rho_{bs}$. This is due to  the fact that, since on each interface side    the bound state wavefunction is a linear combination of two  elementary spinorial waves [see Eq.(\ref{eigen-homo})], $\rho_{bs}$ and $s_{y,bs}$ are determined by different combinations of $w$-spinor components of the wavefunctions, resulting also into different weights for the space-dependent profiles.

\subsection{Effects of a smoothened profile and parallel magnetic field on the orthogonal spin density}
\begin{figure} 
\centering
\includegraphics[width=\columnwidth]{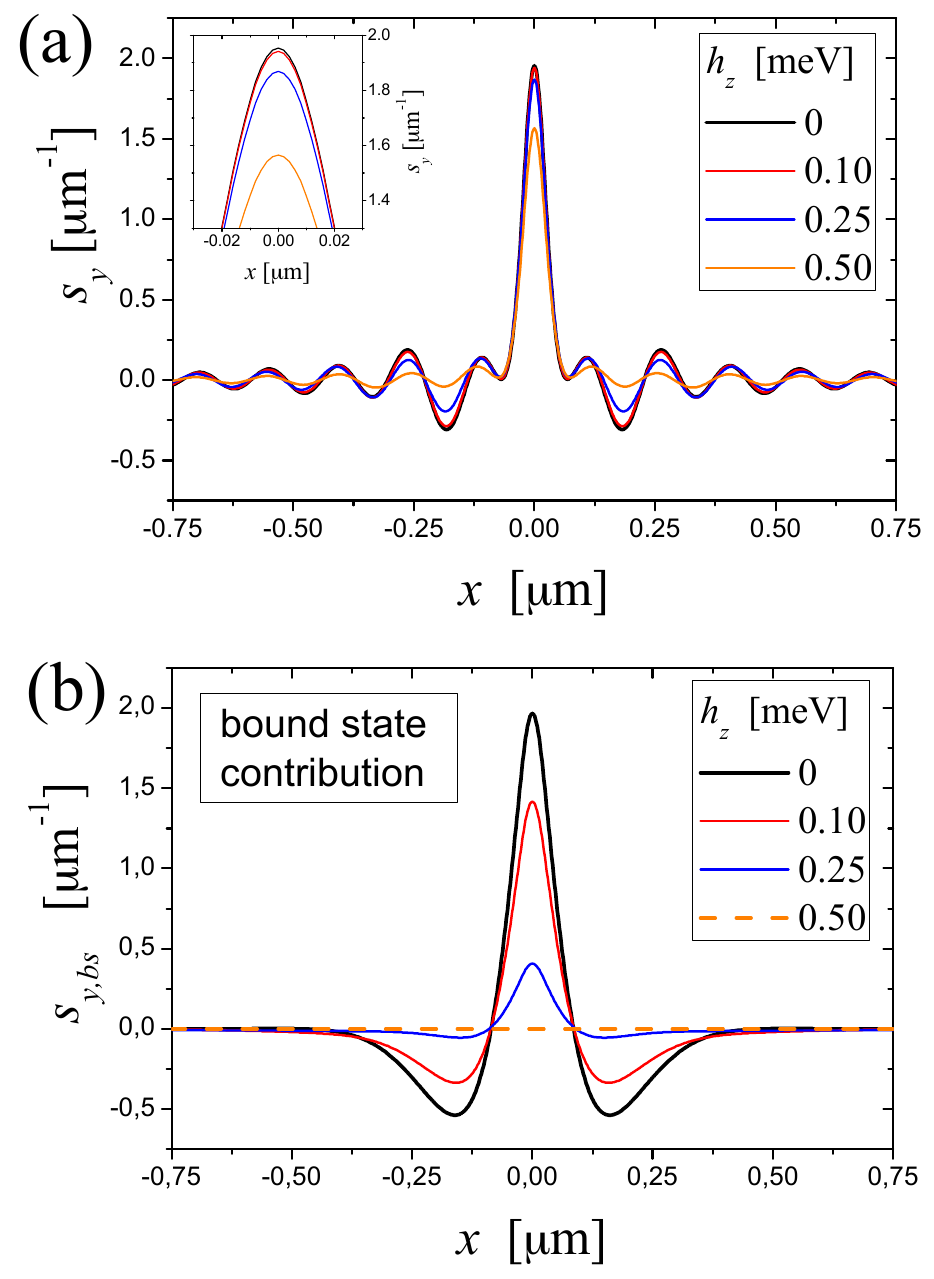}\\
\caption{(Color online) Spatial profile of the orthogonal spin density for a NW interface with $\alpha_L/\alpha_R=-1$ and a smoothening length of $\lambda_s=50\,{\rm nm}$. The maximal spin-orbit energy is $E_{SO,R}=0.25 \,{\rm meV}$, and the magnetic gap is $\Delta_Z=0.50\,{\rm meV}$. Different curves refer to different values of the magnetic field component $h_z$ parallel to the spin-orbit field. (a) The actual~$s_y$ due to all states,  with the inset magnifying the peaks. (b) The bound state contribution to $s_y$.   }
\label{Fig5}
\end{figure}
In the previous subsection we have shown that  the peak of the orthogonal spin density is far more robust than the bound state. In order to test how general such effect is,  we now extend the previous analysis including the presence  of a finite smoothening length in the RSOC profile and  a magnetic field component $h_z$ parallel to the spin-orbit field. For simplicity, we focus on the optimal configuration  $\alpha_L/\alpha_R=-1$ and $E_{SO,R}=0.25 \,{\rm meV}$, with a smoothening length $\lambda_s=50\,{\rm nm}$. These are the parameters also used 
in Fig.\ref{Fig3}(b), whence we observe that, keeping a fixed value of the magnetic gap energy $\Delta_Z$, and varying the additional parallel field component $h_z$ represents a natural physical knob to control the weight and the existence of the bound state. 

Figure~\ref{Fig5} shows the spatial profile of the orthogonal spin density for   $\Delta_Z=0.50\,{\rm meV}$ and  for various values of $h_z$. In particular, panel (a) displays  the full  $s_y$, while   panel (b) shows   the bound state contribution   $s_{y,bs}$.
Two features are noteworthy. In the first instance, as compared to the cuspid peaks obtained at the  interface in the case of the sharp profile [Fig.\ref{Fig4}(c)-(d)], the peaks of Fig.\ref{Fig5} are rounded off by the finite smoothening length~$\lambda_s$. Secondly, while the peak of the full $s_y$ [Fig.\ref{Fig5}(a)] is very weakly affected by the parallel magnetic field component $h_z$,  the bound state peak shown in Fig.\ref{Fig5}(b) rapidly decreases and eventually disappears when the parallel magnetic field component $h_z$ is ramped up, yielding a vanishing contribution (dashed line). This is in agreement   with the binding energy behavior previously shown in Fig.\ref{Fig3}(b), where one can see that, at $\Delta_Z=0.50\,{\rm meV}$, the bound state   disappears for $h_z=0.50\,{\rm meV}$.   The comparison between panels (a) and (b) of Fig.\ref{Fig5} clearly indicates that, when the bound state exists and has a relatively high binding energy, the peak of $s_y$ is mainly due to it. However, when the  binding energy decreases, the bound state contribution to the peak is replaced by the one of the excited states, so that the orthogonal spin density peak remains present.

\section{Discussion}
\label{sec-5}
We have demonstrated that the peak of the orthogonal spin density localized at the interface does not necessarily stem from a localized bound state, and appears to be a quite general feature. Two natural questions then arise, namely  i) what  parameters characterizing the interface determine such peak? ii) can one explain its presence on some general principle? Here we wish to address these two questions. 
\subsection{General features of the orthogonal spin density}
\label{sec-5-A}
\begin{figure} 
\centering
\includegraphics[width=\columnwidth]{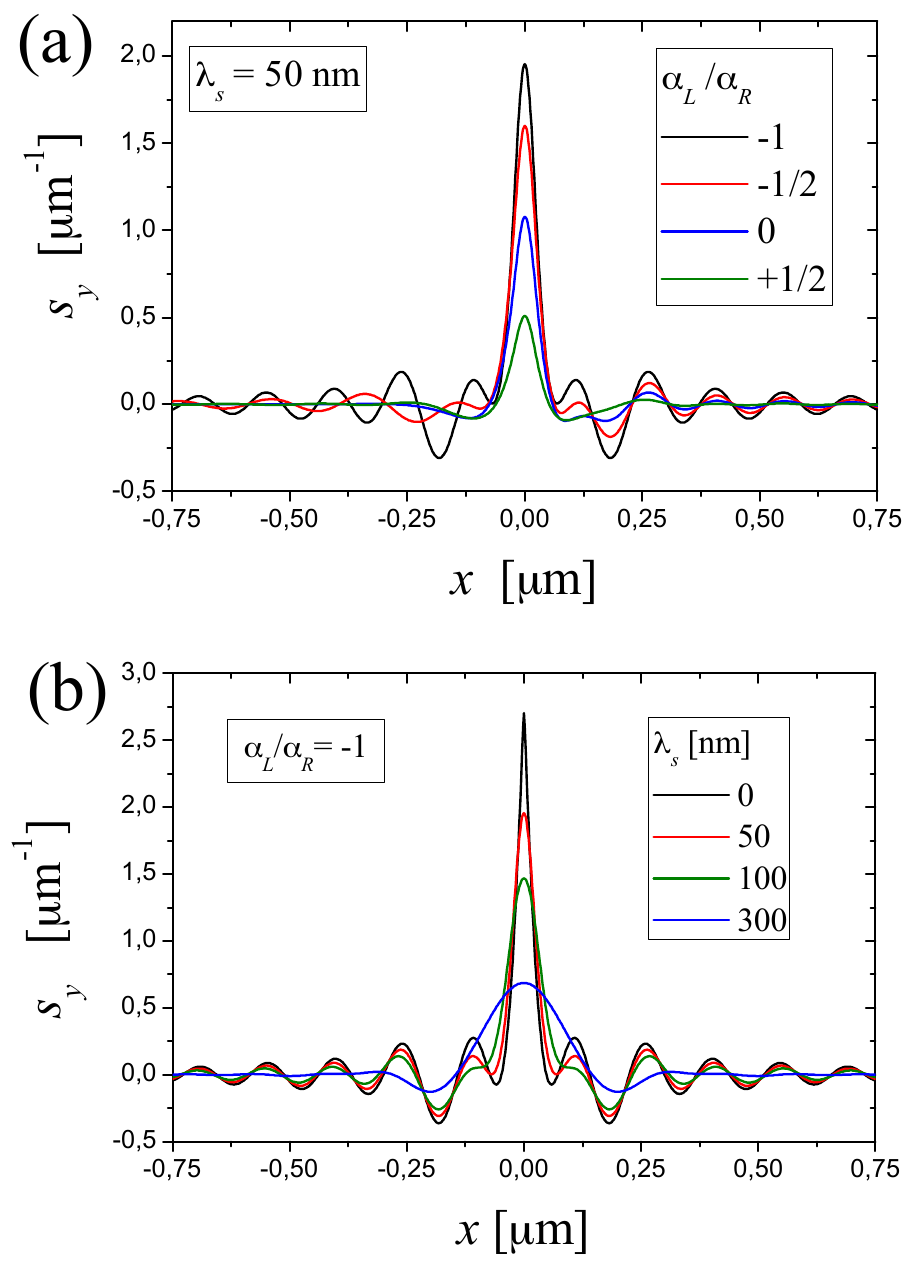}\\
\caption{(Color online) Spatial profile of the orthogonal spin density for an interface with  $E_{SO,R}=0.25\, {\rm meV}$, and a magnetic gap energy $\Delta_Z=0.50\,{\rm meV}$. (a) The effects of the ratio between the two values of RSOC, for a fixed smoothening length $\lambda_s=50\,{\rm nm}$. (b) Effects of the smoothening length, for the configuration $\alpha_L/\alpha_R=-1$.}
\label{Fig6}
\end{figure}

To answer the first question, we consider for definiteness   the case of magnetic gap energy  $\Delta_Z=0.50\,{\rm meV}$ and a 
 maximal spin-orbit energy  $E_{SO,R}=0.25\, {\rm meV}$. Two parameters characterize the interface, namely the ratio $\alpha_L/\alpha_R$ of the two RSOC, and the smoothening length of the profile.
In Fig.\ref{Fig6}(a) we show, for a fixed  smoothening length   $\lambda_s=50\,{\rm nm}$,   the orthogonal spin density profile  for different  values of the RSOC ratio $\alpha_L/\alpha_R$  across the interface.  As one can see, the height of the peak grows with the relative RSOC jump, in a roughly linear way.  \\
In Fig.\ref{Fig6}(b), keeping now the ratio of the two RSOC  bulk values to $\alpha_L/\alpha_R=-1$, we vary the smoothening length $\lambda_s$ of the profile.  The peak decreases and broadens with increasing $\lambda_s$. Importantly, one can verify by a numerical integration that the area underneath each $s_y(x)$ profile is to a very good approximation {\it independent} of the value of the smoothening length $\lambda_s$.

\subsection{Origin of the orthogonal spin density}
\label{Sec-5-B}
Keeping in mind the  two features described in the previous subsection, let us now discuss the origin of the orthogonal spin density peak.   
As is well know,  a magnetic moment exposed to a magnetic field experiences a magnetic torque~[\onlinecite{sonin_2010}]. So is the case for spin magnetic moments of electrons moving in a NW, where  both the  externally applied magnetic field $\mathbf{h}$ and the effective spin-orbit field $\mathbf{h}^{SO}$ give rise to corresponding torques, defined as 
\begin{eqnarray}
\hat{\mathbf{T}}^{h} &\equiv &    \hat{\Psi}^\dagger       \left(   \boldsymbol\sigma \,  \times \mathbf{h}\right) \hat{\Psi}\quad,\label{Th-def} \\
\hat{\mathbf{T}}^{SO} & \equiv &  \displaystyle \frac{1}{2} \left( \hat{\Psi}^\dagger  ( \boldsymbol\sigma \,  \times  \mathbf{h}^{SO} )   \hat{\Psi}^{}+{\rm H.c.}  \right)\,   \quad,  \label{TSO-def} 
\end{eqnarray}
respectively,  where
\begin{equation}\label{SO-field-op}
\mathbf{h}^{SO}(x,t)=\frac{\left\{\alpha(x),p_x\right\}}{2\hbar}(0,0,1)
\end{equation}
is the spin-orbit field.
Note that, by definition Eqs.(\ref{TSO-def})-(\ref{SO-field-op}), the spin-orbit torque $\hat{\mathbf{T}}^{SO}=(\hat{\rm T}^{SO}_x,\hat{\rm T}^{SO}_y,0)$ has no component along the Rashba field direction $z$. \\

Importantly, the torques determine the spin-dynamics through the operator identity
\begin{equation}\label{spin-cont-eq}
\partial_t \hat{\mathbf{S}} +\partial_x \hat{\mathbf{J}}^s=\, \hat{\mathbf{T}}^{h}+\, \hat{\mathbf{T}}^{SO}  
\end{equation}
where $\hat{\mathbf{S}}$ is the spin density operator in Eq.(\ref{spin-density-def}), and 
\begin{eqnarray}
\hat{\mathbf{J}}^s\!  
&=& \frac{\hbar}{2} \,\left(-\frac{i\hbar}{2 m^*} \left( \hat{\Psi}^\dagger(x) \, \boldsymbol\sigma\, \partial_x\hat{\Psi}^{}(x) -\partial_x \hat{\Psi}^\dagger(x) \, \boldsymbol\sigma\, \hat{\Psi}^{}(x) \right) \,\, \right.\nonumber \\
& & \left. \hspace{1cm} -\frac{\alpha(x)}{\hbar} \hat{\Psi}^\dagger(x) \, \frac{\{ \boldsymbol\sigma, \sigma_z\}}{2} \, \hat{\Psi}^{}(x) \right) \quad  \label{spin-current-def}
\end{eqnarray}
is the spin current density operator~[\onlinecite{sonin_2010},\onlinecite{rashba_2003}]. Differently from the continuity equation for charge, in Eq.(\ref{spin-cont-eq}) the torques on the right-hand side play the role of sources and sinks of spin.

At equilibrium  the  expectation values of $\hat{\mathbf{S}}$ is time-independent, while the one of the magnetic torque  is straightforwardly related to the equilibrium spin-density  Eq.(\ref{s-def}), through  ${\rm\mathbf{T}}^{h} =\langle \hat{\mathbf{T}}^{h} \rangle_\circ= \mathbf{s}  \times \mathbf{h}$. Thus, taking the equilibrium expectation value Eq.(\ref{spin-cont-eq}) one has
\begin{equation} \label{cont-exp-pre}
\partial_x \mathbf{J}^s= \mathbf{s}(x) \times \mathbf{h} + {\rm\mathbf{T}}^{SO}(x)
\end{equation}
where ${\rm\mathbf{T}}^{SO} =\langle \hat{\mathbf{T}}^{SO} \rangle_\circ$. Let us focus on the most customary situation where the magnetic field is directed along the NW axis $x$ ($\mathbf{h}=h_x \mathbf{i}_x$),  i.e., orthogonal to the spin-orbit field. In this case, one can show that the spin-orbit torque ${\rm\mathbf{T}}^{SO}(x)$ vanishes, and that the spin current is oriented along $z$, so that 
Eq.(\ref{cont-exp-pre}) reduces to 
\begin{equation} \label{cont-exp}
\partial_x J^s_z= - h_x  \, s_y(x)   \quad.
\end{equation}
We shall now argue that this equation, derived under quite general hypotheses,  is the key to interpret the appearance of the orthogonal spin density at the interface, even when the bound state is absent.\\

Indeed, as has been demonstrated in Ref.[\onlinecite{dolcini-rossi_PRB_2018}], when   uniform spin-orbit and    magnetic fields  are present in a NW, an equilibrium spin current $J^s_z$ flows in its bulk. Such bulk spin current arises from  the interplay between spin-orbit field and a magnetic field orthogonal to it, which induce non-trivial quantum correlation between spin and velocity, in close similarity to what happens in the helical states of a quantum spin Hall system.  The bulk equilibrium spin current  is odd in $\alpha$ and even in $h_x$. For example, for $\mu=0$ and in the regime $\Delta_Z \gg E_{SO}$,   one has $J^s_z=-\mbox{sgn}(\alpha) \sqrt{\Delta_Z E_{SO}}/3\pi$.  
Equilibrium spin currents have been predicted for other RSOC systems as well~[\onlinecite{rashba_2003,governale_2003,balseiro_2005,sonin_PRB_2007,sonin_PRL_2007,wang_PRL_2007,wang_PRB_2008,sablikov_2008,liang_PLA_2008,medina_2010,sonin_2010,nakhmedov_2012,
liu_2014,liang_PLA_2015,chen_PRB_2006,loss-meier_2002}] and, in fact, they can be regarded to as the diamagnetic color currents associated to the non-abelian spin-orbit gauge fields~[\onlinecite{tokatly_2008}]. However, 
its measurement in actual experiments has not been achieved thus far. In this respect, Eq.(\ref{cont-exp}) suggests that, while the equilibrium spin current itself is perhaps elusive, its {\it variation} in the presence of inhomogeneities could be detected, as it is straightforwardly connected to the orthogonal spin density. Indeed, when two regions with different RSOC are connected, a kink $\partial_x J^s_z$ must arise at the interface to match the different  spin current values in the two bulks. In view of  Eq.(\ref{cont-exp}), a peak in the orthogonal spin density $s_y$ necessarily appears. 
This is the reason why the peak of $s_y$ shown in Fig.\ref{Fig6}(a) is the more pronounced  the higher the difference in the RSOC of the two regions. Furthermore, integrating both sides of  Eq.(\ref{cont-exp}), one can see that the integral of the $s_y$  profile equals the difference between the two bulk spin currents, which is  independent of the smoothening length. This is precisely what we found in Fig.\ref{Fig6}(b).
Finally,  this argument is quite general and is not based on the existence of a bound state at the interface. This explains why the peak shown in Fig.\ref{Fig4}(d) persists even when the bound state is absent, and shows that the naive interpretation of an orthogonal spin density localized peak in terms of a bound state is in general wrong.

\subsection{The case of two interfaces} 
\begin{figure} 
\centering
\includegraphics[width=\columnwidth]{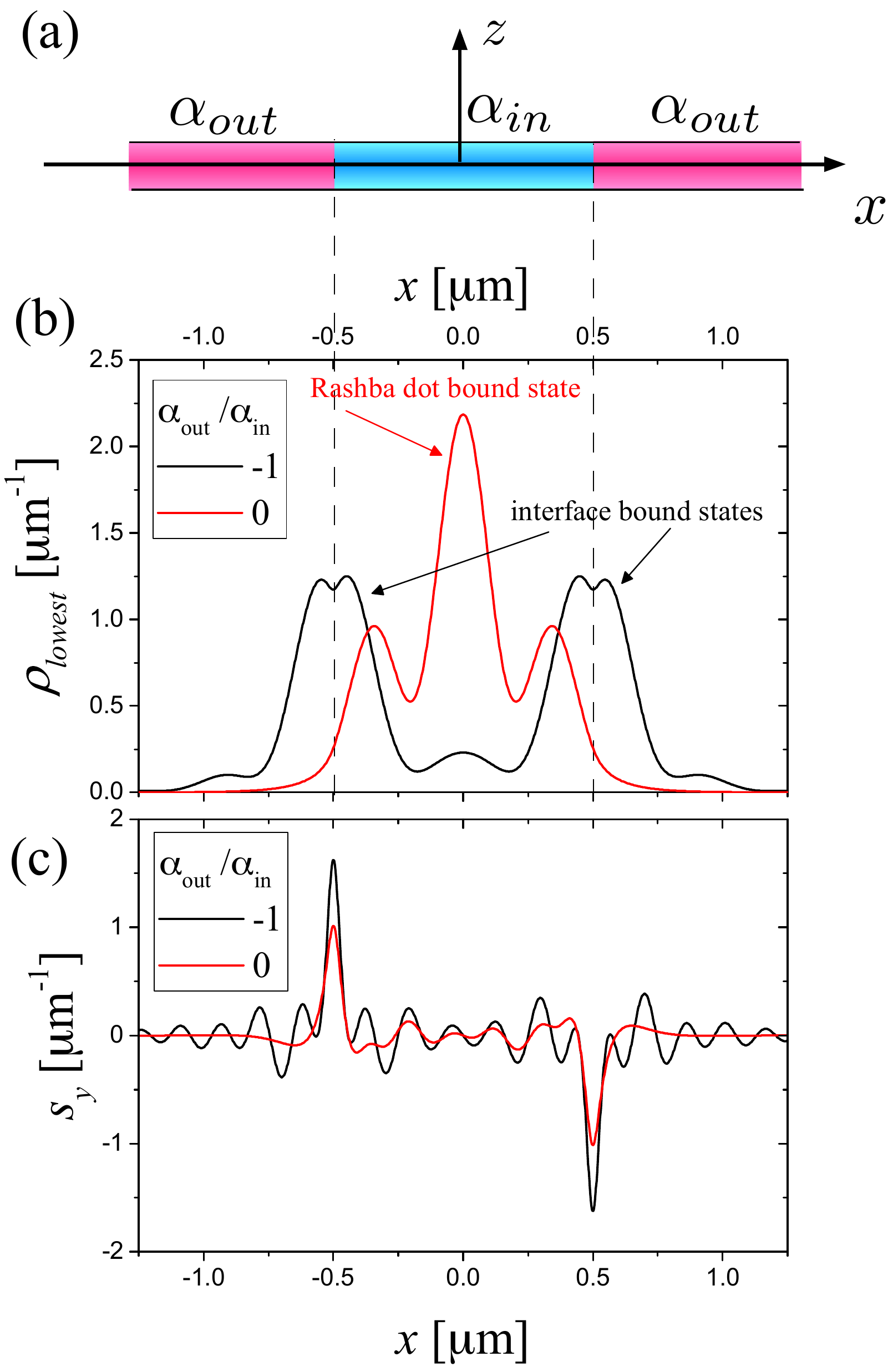}\\
\caption{(Color online)  (a) Sketch of a double interface problem, modeled by the RSOC profile (\ref{alpha-double}). The parameters are $L=1\,\mu {\rm m}$, $\lambda_s=50\,{\rm nm}$,  the value $\alpha_{in}>0$ in the inner region corresponds to $E_{SO,in}=0.25\, {\rm meV}$, while the magnetic gap energy is $\Delta_Z=0.25\,{\rm meV}$. (b) The density profile of the lowest electron state,  for two values $\alpha_{out}=-\alpha_{in}$ (black curve) and $\alpha_{out}=0$ (red curve), showing the difference between interface and Rashba dot bound states. (c) The total orthogonal spin density, for the same two values of $\alpha_{in}$, shows two opposite peaks at the interfaces.  }
\label{Fig7}
\end{figure}

Thus far, we have considered the case of one single interface along the NW. Here we wish to briefly discuss the case of two interfaces,   modeling a NW inner region characterized by a RSOC parameter $\alpha_{in}$ sandwiched between two outer regions, where  the RSOC shall be taken  for simplicity equal to $\alpha_{out}$ in both. This corresponds to a profile
\begin{eqnarray}
\alpha(x)&=& \alpha_{out}+\frac{\alpha_{in}-\alpha_{out}}{2} \label{alpha-double}\\
& & \times \left[{\rm Erf}\left(\frac{\sqrt{8}}{\lambda_s}(x+\frac{L}{2}) \right) -{\rm Erf}\left(\frac{\sqrt{8}}{\lambda_s}(x-\frac{L}{2})\right) \right] \quad,\nonumber
\end{eqnarray}
sketched in Fig.\ref{Fig7}(a), where $L$ denotes the length of the inner NW region, supposed to be much bigger than the smoothening length ($L \gg \lambda_s$), so that the notion of interfaces still makes sense. When the distance $L$ is much larger than the typical variation lengthscale for observables in the single interface problem, the two interfaces act independently.  
However, when such two scales become comparable, noteworthy aspects emerge, which are illustrated in Fig.\ref{Fig7}. 

First, if the   interface bound states exist, they  overlap across the   distance $L$, causing a splitting of their degeneracy. The density profile of the resulting lowest eigenstate is mainly peaked at the interfaces, but  is non vanishing also in the center of the inner region, as illustrated by the black curve in  Fig.\ref{Fig7}(b). Second, even if the interface bound states are not present, another type of bound states may appear. Indeed, when the inner region is Rashba-dominated and   $|\alpha_{in}|>|\alpha_{out}|$, the band bottom of the inner region is lower than in the outer regions. Thus, for short $L$ the two interfaces  give rise to an  effective   Rashba quantum dot[\onlinecite{sanchez_2006},\onlinecite{sanchez_2008}], with discrete bound states  localized within the confinement length $L$. This is the case depicted by the red curve  in  Fig.\ref{Fig7}(b), where the density profile of the lowest eigenstate corresponds to a Rashba dot bound state.   
Such quantum dot bound states thus have a completely different origin from the interface bound states. In particular, while the interface bound states are present only in the presence of an applied magnetic field, the Rashba dot bound states are intrinsic, as they may also be present without magnetic field[\onlinecite{nota-gauge}].

 The third interesting feature of the double interface problem is that, in all cases,  pronounced orthogonal spin density peaks appear  at the interfaces, regardless of whether interface bound states exist or not. 
Remarkably, the signs of the peaks are {\it opposite} at the two interfaces, as shown in Fig.\ref{Fig7}(c). This is because the opposite jump in the RSOC  across the two interfaces causes two opposite kinks in the equilibrium spin current, as observed in Sec.\ref{Sec-5-B}. 
Thus, despite the NW  is in the topologically trivial phase, the emerging scenario is identical to the one occurring in a NW in the topological phase, where the  spin density of the MQPs is orthogonal to both the magnetic field and the RSOC field direction, and takes  opposite signs at the two NW ends[\onlinecite{simon_PRL_2012,black-schaffer_2015,domanski_scirep_2017}]. This explicitly demonstrates that such orthogonal spin polarization pinned at the NW ends can neither be taken as a hallmark of the topological phase, nor as an evidence of  bound states.   Note also that the orthogonal spin polarization peaks are typically narrower than the interface bound state and are thus more robust to   finite length $L$ effects too.

\subsection{Possible setup realizations} 
Several experiments in topological systems are based on
InSb~[\onlinecite{xu_2012,nilsson_2009,kouwenhoven_PRL_2012,wimmer_2015,kouwenhoven_nanolett_2017,kouwenhoven_2018}] or InAs~[\onlinecite{ensslin_2010, gao_2012,joyce_2013,nygaard_2016,soc4,marcus_science_2016}] NWs deposited on a substrate.
 In the  case of InSb the effective mass and the  $g$-factor are  $m^* \simeq 0.015 m_e$ and $g\simeq 50$, respectively, while the value of the RSOC depends on the specific implementation and experimental conditions and  can be widely tunable,  ranging from $\alpha \sim  0.03  \, {\rm eV} \, {\rm \AA}$ to $\alpha \sim  1 \, {\rm eV} \, {\rm \AA}$~[\onlinecite{kouwenhoven_2012},\onlinecite{xu_2012},\onlinecite{gao_2012},\onlinecite{wimmer_2015},\onlinecite{nilsson_2009},\onlinecite{kouwenhoven_PRL_2012}]. The spin-orbit energy $E_{SO}$ resulting from these values  [see Eq.(\ref{ESO-def})] is a fraction of ${\rm meV}$.  The same order of magnitude is obtained for the magnetic gap energy~$\Delta_Z$ in a magnetic field range of  some hundreds of ${\rm mT}$. These are the values adopted in our plots. 
Similarly, in the case of InAs nanowires $m^*\simeq 0.022\,m_e$, $g\simeq 20$ and  the RSOC ranges from $\alpha \sim  0.05  \, {\rm eV \, \AA}$ to $\alpha \sim  0.3  \, {\rm eV \, \AA}$~[\onlinecite{heiblum_2012},\onlinecite{ensslin_2010},\onlinecite{gao_2012},\onlinecite{joyce_2013}]. 
 The temperature value of $250\,{\rm mK}$ used in our plots  is state of the art with modern refrigeration techniques.  
 
Interfaces between regions with different RSOC emerge quite naturally in typical NW setups, where a portion of the NW is covered by   e.g.  a superconductor or by a normal metal  to induce proximity effect,  to measure the current, or to locally vary the potential. The resulting SIA is  inhomogeneous along the NW, and can be controlled e.g. by the application of   different gate voltage values applied to   top/bottom gates or to the substrate, similarly to the case of constrictions in quantum spin Hall systems[\onlinecite{sassetti-citro_2012,sternativo_2014,molenkamp-buhmann_2020}].
In particular, covering one portion with the gate-all-around technique and by applying a sufficiently strong gate voltage, it is reasonable to achieve an inversion of the sign of the RSOC as compared to the uncovered NW portion, as has already been done in similar setups[\onlinecite{slomski_NJP_2013,kaindl_2005,wang-fu_2016,nitta-frustaglia,tsai_2018}].

Finally, the orthogonal spin polarization predicted here can be measured by spatially resolved  detection of spin orientation. In particular, nanometer scale resolution can be reached with various methods such as magnetic resonance force microscopy~[\onlinecite{chui_2004},\onlinecite{hammel_2015}],   spin-polarized scanning electron microscopy~[\onlinecite{koike_1985},\onlinecite{kohashi_2015}], by using quantum dots as probes~[\onlinecite{katsumoto_2009},\onlinecite{tarucha_2012}], or also electrically by potentiometric measurements exploiting ferromagnetic detector contacts~[\onlinecite{jonker_2014},\onlinecite{wang_2014}]. 

\section{Conclusions}
\label{sec-6}
In conclusions, in this paper we have considered a NW with an interface between two regions with   different RSOC values, as sketched in Fig.\ref{Fig1}, when proximity effect is turned off and the NW is in the topologically trivial phase.

In Sec.\ref{sec-3} we  have shown that  at the interface bound states may appear, whose energy is located below the continuum spectrum minimum. Such bound states are neither topological (since proximity effect is absent), nor intrinsic interface bound states (since they only exist if an external magnetic field is applied along the NW axis). Analyzing first the case of a sharp interface RSOC profile Eq.(\ref{alpha-sharp}), we have obtained the phase diagram determining the existence of the bound state [see Fig.\ref{Fig2}(a)], as well as  the dependence of its binding energy on the magnetic gap energy [see Fig.\ref{Fig2}(b)]. While the bound state always exists if the RSOC takes equal and opposite values across the interface (optimal configuration), for all other situations it only exists if the magnetic field overcomes a minimal threshold value.   Furthermore,   even in the optimal configuration, it can be suppressed  by either a finite smoothening length in the RSOC profile or a magnetic field component parallel to the spin-orbit field  (see Fig.\ref{Fig3}). \\

In Sec.\ref{sec-4} we have then investigated the spatial profile of the charge density $\rho$ and the spin density, with a special focus on the spin   density component $s_y$, orthogonal   to both the applied magnetic field  and the RSOC field direction, which is known to characterize the MQPs localized at the  edges of a NW in the topological phase.   By analyzing both  the full equilibrium values $\rho$ and $s_y$ due to all occupied states, and  the bound state contributions  $\rho_{bs}$ and $s_{y,bs}$, we have been able to gain two useful insights. First, the   orthogonal spin density  appears also in the topologically trivial phase as a quite general effect characterizing any interface between two different RSOC regions under a magnetic field. This extends  our previous results of Ref.[\onlinecite{dolcini-rossi_PRB_2018}] related to NW contacted to normal leads without RSOC. Second,  for realistic and typical values of chemical potential and temperature,  the orthogonal spin density peak is relatively robust to parameter changes, and persists even when the bound state is absent  (see Figs.\ref{Fig4} and \ref{Fig5}).  This means that also the propagating states of the continuum spectrum modify their spin texture around the interface to preserve the   peak, so that a localized orthogonal spin-density cannot   be considered a signature of a bound state.   

Furthermore, in Sec.\ref{sec-5}, after analyzing in Fig.\ref{Fig6} the peak  dependence on  the  single interface parameters, we have addressed the case of two interfaces  [see Fig.\ref{Fig7}]. While for a large distance $L$ between the interfaces the single-interface scenario is merely doubled, for a shorter~$L$ the interface bound states may overlap and additional Rashba quantum dot states may appear. In all cases, and independently of the presence of interface bound states, the spin density   $s_y$,  {\it orthogonal}  to both the magnetic field and the Rashba spin-orbit field,  exhibits  relatively robust  peaks   taking {\it  opposite  signs}  at the two interfaces  [see Fig.\ref{Fig7}(c)]. Remarkably, these are   the same features predicted for the spin density of the MQPs emerging at the ends of a NW in the topological phase, despite the NW considered here  is  in the topologically trivial phase. Our results thus show  that such orthogonal spin polarization pinned at the NW ends can neither be taken as a hallmark of the topological phase, nor as an evidence of  bound states.   

However, we have also shown in Sec.\ref{sec-5} that   such stable peaks may in fact have an impact on the detection of spin currents. Indeed a spin current flows in the bulk of a NW as a result of  quantum correlations between spin and velocity induced by the interplay between  magnetic and spin-orbit field,  similarly to    the case of quantum spin Hall  helical states.  Despite various proposals in the literature, the measurement  of   equilibrium spin currents has not been achieved yet. Our results suggest that, while the equilibrium spin current itself may be elusive, its variations can be detected  through the orthogonal spin density $s_y$, which is instead experimentally observable with spin-resolved detection techniques. Indeed the orthogonal spin density peak is precisely related to the  kink of the spin current localized at the interface.  
 With the provided description of possible implementations  in realistic NW setups, the predicted effects seem to be at experimental reach.

\acknowledgments

Fruitful discussions with M. Sassetti, F. Cavaliere, and N. Traverso Ziani are greatly acknowledged. Computational resources were provided by \href{http://www.hpc.polito.it}{hpc@polito} at Politecnico di Torino.\\

\appendix
\section{Calculation for sharp profile interface}
\label{AppA}
 In this Appendix we provide details about the calculation for a sharp profile interface (\ref{alpha-sharp}). In such a situation the eigenvalue equation stemming from the Hamiltonian (\ref{H-realspace}) at energy $E$ reads
\begin{widetext}
\begin{equation}\label{sharp eigenvalue equation}
\begin{pmatrix}
    -\frac{\hbar^2}{2 m^*} \partial_x^2+i \alpha(x) \partial_x +i \frac{\alpha_R-\alpha_L}{2} \delta (x)-h_z& -h_x \\
   -h_x & -\frac{\hbar^2}{2 m^*} \partial_x^2-i \alpha(x) \partial_x -i \frac{\alpha_R-\alpha_L}{2}\delta (x)+h_z
\end{pmatrix}
\begin{pmatrix}
   \psi_\uparrow^{(E)}(x) \\
   \psi_\downarrow^{(E)} (x)
\end{pmatrix}
= E
\begin{pmatrix}
   \psi_\uparrow^{(E)}(x) \\
   \psi_\downarrow^{(E)} (x)
\end{pmatrix}
\end{equation}

\end{widetext}
equipped with the boundary conditions at the interface 
\begin{equation}\label{boundary conditions}
\begin{cases}
\psi_\uparrow (0^-) = \psi_\uparrow (0^+)\\
\psi_\downarrow (0^-) = \psi_\downarrow (0^+)\\
\partial_x \psi_\uparrow (0^-) = \partial_x \psi_\uparrow (0^+) -i\frac{m^*}{\hbar^2} ( \alpha_R-\alpha_L) \psi_\uparrow (0)\\
\partial_x \psi_\downarrow (0^-) = \partial_x \psi_\downarrow (0^+)+i \frac{m^*}{\hbar^2} ( \alpha_R-\alpha_L) \psi_\downarrow (0)\\
\end{cases}
\end{equation}
A few remarks   about   the boundary conditions (\ref{boundary conditions}) are in order. First, the discontinuity in the derivative of the wavefunction  involves an imaginary unit too, making such boundary conditions intrinsically different from the ones of the well known problem of a particle in a scalar $\delta$-potential. Second, as a consequence of such imaginary unit, it can straightforwardly be shown that, despite the derivative $\partial_x \psi_s$ is discontinuous  ($s=\uparrow,\downarrow$),  the derivative $\partial_x\rho_s$ of the quantity $\rho_s(x)\equiv \psi^*_s(x)\psi^{}_s(x)$ is {\it continuous} at the interface $x=0$. For this reason, both the density $\rho(x)=\rho_\uparrow+\rho_\downarrow$ [see Eq.(\ref{rho-def})] and the spin density component $s_z=\rho_\uparrow-\rho_\downarrow$ [see Eq.(\ref{s-def})] do not exhibit any cusp in their spatial profile. In contrast, off-diagonal spin density components $s_x$ and $s_y$, which cannot be expressed in terms of the $\rho_s$'s, do exhibit a cusp   due to the discontinuity of the derivative implied by  the boundary conditions (\ref{boundary conditions}). This difference becomes apparent by comparing e.g. panels (b) and (d) in Fig.\ref{Fig4}.\\

Let us now proceed with the calculation of the energy spectrum. As observed above,  we have assumed $\alpha_R>0$ and $|\alpha_L|\le |\alpha_R|$ without loss of generality. As a consequence $E_{SO,R}$ is the higher spin-orbit energy, $E_{SO,R} \ge E_{SO,L}$ [see Eq.(\ref{ESO-nu-def})]. 
By denoting the ratio between the two RSOC values
\begin{equation} \label{r-def}
{r} \equiv \frac{\alpha_L}{\alpha_R} \,\,  \, \in [\, -1 \, , \,1 \,]
\end{equation}
one has $E_{SO,L}= {r}^2 E_{SO,R}$. 
One can introduce the momentum space Hamiltonian $H_k^\nu =\varepsilon_k^0-\alpha_\nu k \sigma_z -h_x \sigma_x-h_z \sigma_z$ describing the homogeneous bulk of each side $\nu=R/L$ of the interface, and match the related eigenfunctions with the boundary conditions (\ref{boundary conditions}). 

The energy spectrum characterizing the NW on the right-hand side and on the left-hand side of the interface can be suitably rewritten as
\begin{eqnarray}
E_{\pm}^{R}(K)&=& \frac{K^2}{4E_{SO,R}} \pm \sqrt{\Delta_Z^2 + (K+h_z)^2} \label{EpmR} \\
E_{\pm}^{L}(K) &=&\frac{({r} K)^2}{4E_{SO,L}} \pm \sqrt{\Delta_Z^2 + ({r} K+h_z)^2}  \label{EpmL}
\end{eqnarray}
respectively, where $K=\alpha_R k $ has the dimension of an energy, while $\Delta_Z$ is the magnetic gap energy  Eq.(\ref{EZ-def}).

The eigenstates of the momentum Hamiltonian in each side can be written, for arbitrary  complex wavevector $K$, in the following explicit form 
\begin{eqnarray}
&\mbox{for } x>0\,\, & \left\{
\begin{array}{ll}
w_{-} (K) &= \frac{1}{\sqrt{\Delta_Z^2+|z(K)|^2}}  \left( \begin{array}{c}  z(K) \\
\Delta_Z \end{array} \right)  \\ &
\\
w_{+}(K)&= \frac{1}{\sqrt{\Delta_Z^2+|z(K)|^2}} \left( \begin{array}{c} -\Delta_Z \\
z(K)\end{array} \right) 
\end{array}
\right.  \label{wpmx>0}
\\
&\mbox{for } x<0 \,\, & \left\{
\begin{array}{ll}
w_{-} ({r} K)&= \frac{1}{\sqrt{\Delta_Z^2+|z({r} K)|^2}} \left( \begin{array}{c}  z({r} K) \\
\Delta_Z\end{array} \right) \\ &
\\
w_{+} ({r} K)&= \frac{1}{\sqrt{\Delta_Z^2+|z({r} K)|^2}}  \left( \begin{array}{c}  -\Delta_Z \\
z({r} K) \end{array} \right) 
\end{array}
\right.  \label{wpmx<0}
\end{eqnarray}
where $z(K)=\sqrt{\Delta_Z^2+(K+h_z)^2}+(K+h_z)$.\\

In order to determine the energy $E_{bs}$ of the bound state, the crucial point is to correctly re-express Eqs.(\ref{wpmx>0})-(\ref{wpmx<0}) as a function of the energy $E$, and then to impose the boundary conditions (\ref{boundary conditions}). 
To this purpose, the first step is to invert the dispersion  relation in each side $\nu=R/L$.
This can be done analytically in two specific cases, namely for $h_z=0$ or for $h_x=0$. Here below we   shall discuss these two  situations, while the general case $h_x, h_z \neq 0$ will be approached numerically as described in App.\ref{AppB}.\\

\subsection{The case $h_z=0$} 
In this case the dispersion relation can be inverted yielding four possible  $K$-values
\begin{eqnarray}
\lefteqn{K_{\epsilon, \epsilon^\prime}^{\nu} (E) = } & & \label{Knu} \\
& &\epsilon \sqrt{4E_{SO,R}\left[E + 2E_{SO,\nu} +\epsilon^\prime \sqrt{\Delta_Z^2 +4E_{SO,\nu}^2+4E_{SO,\nu}E} \right]}   \nonumber 
\end{eqnarray}
where $\epsilon, \epsilon^\prime = \pm 1$. Note that $K \in \mathbb{C}$, and  we have adopted the   convention 
 $ \sqrt{z}=\sqrt{|z|} e^{i \frac{\phi}{2}}
$ for the square root of a complex number $z=|z| e^{i \phi} $ with $\phi \in \, ( - \pi \, , \, \pi \,]$. 

One then inserts the four possible values (\ref{Knu}) of $K_{\epsilon, \epsilon^\prime}^{\nu}$ into  the two eigenvectors Eqs.(\ref{wpmx>0})-(\ref{wpmx<0}). In doing that, some caution must be taken, since for a given energy $E$ and   each side of the interface  a seeming redundancy of eigenstates appears. However, only half of the possible eigenstates actually fulfill the equation $H_k \left[ K(E) \right] w\! \left[ K(E) \right]=E \, w\! \left[K(E) \right]$, as it should be. Their explicit expressions depend on the regime of  the  involved energy scales $E$, $\Delta_Z$ and $E_{SO,\nu}$,
Focusing e.g. on the right hand side of the interface, one can identify three regimes  where, for a given energy $E$ lower than the overall   minimum of the bulk bands, the corresponding 4 correct eigenspinors are given in Table Eq.(\ref{table-eigenv}).

Regime 2 differs from regime 3 because in the former wave vectors turn out to be strictly imaginary, while in the latter they exhibit a real part as well. 
The   expression for the eigenspinors on the left hand side, together with their corresponding domain, can be directly obtained from the  ones in Table~(\ref{table-eigenv}) by simply replacing 
$
E_{SO,R} \rightarrow E_{SO,L}$ and $K_{\pm \pm}^{R}(E) \rightarrow {r}\, K_{\pm \pm}^{L}(E)$.\\

\begin{widetext}
\begin{equation}
\begin{tabular}{ |l|c| }\hline
\quad\quad  \mbox{regime}  & \mbox{eigenvectors}  \\ \hline
1) \quad $\Delta_Z > 4E_{SO,R} $ \,\,\mbox{and} \,\,\,  $\displaystyle
-\frac{\Delta_Z^2+4E_{SO,R}^2}{4E_{SO,R}}<E<-\Delta_Z$  & 
$  \begin{array}{l}  
   w_{-}\!\left[ K_{\epsilon, +}^{R}(E) \right] \\
   w_{+}\! \left[ K_{\epsilon, +}^{R}(E) \right] 
   \end{array}$   \quad $\epsilon=\pm 1$
 \\ \hline  
2) \quad $\Delta_Z > 2E_{SO,R}$  \,\,\mbox{and} \,\,\,  $
\displaystyle -\frac{\Delta_Z^2+4E_{SO,R}^2}{4E_{SO,R}}<E<{\rm min}\left[ -\frac{\Delta_Z^2}{4E_{SO,R}},-\Delta_Z \right]$  & $w_{-}\! \left[ K_{\epsilon, \epsilon^\prime}^{R}(E) \right]$  \quad $\epsilon,\epsilon^\prime=\pm 1$  \\ \hline
3) \quad $\Delta_Z <4E_{SO,R}$ \,\,\mbox{and} \,\,\,  $\displaystyle 
E<-\frac{\Delta_Z^2+4E_{SO,R}^2}{4E_{SO,R}}$  & $w_{-}\! \left[ K_{\epsilon, \epsilon^\prime}^{R}(E) \right]$ \quad $\epsilon,\epsilon^\prime=\pm 1$ \\
 \hline 
\end{tabular} \label{table-eigenv}
\end{equation}
\end{widetext}

 Once the four eigenspinors $w$ and momenta $K$ are identified, the wavefunction $\psi$ is constructed as a linear superposition of  each spinor $w$ multiplied by the related phase factor $e^{i K x /\alpha_R}$. In doing that, the requirement that  $\psi$ does not diverge at $x\rightarrow \pm \infty$   reduces the four terms to two in each side. 
Let thus $w^\nu_{j}(E)$ and $K^\nu_{j}(E)$ with $j=1,2$ denote such two eigenspinors and momenta related to non-divergent wavefunctions   in the region $\nu=R/L$ at energy $E$ in a given regime.   
Then, the eigenfunction $\psi^{(E)}(x)$ can be written as a linear superposition   
\begin{equation}
\psi^{(E)}(x)=
\begin{cases}
\sum_{j=1}^2 l_j w^R_{j} (E)\, e^{i \frac{K^R_{j}(E) }{\alpha_R} x} \qquad &x>0 \\
\sum_{j=1}^2 r_j w^L_{j} (E)\, e^{i \frac{ K^L_{j}(E) }{\alpha_R} x} \qquad & x<0
\end{cases} \quad.
\end{equation}
Thus,  the boundary condition Eq.(\ref{boundary conditions}) leads to a homogeneous system of 4 linear equations in 4 unknowns $l_{1}$, $l_{2}$ $r_{1}$ and $r_{2}$. Imposing the solvability of the system one obtains an equation for the energy $E$ whose solutions, if they exist, correspond to the energy  $E_b$ of the bound state for given values of $\Delta_Z$, $E_{SO,R}$ and~${r}$. The binding energy (\ref{Eb-def}) is then straightforwardly obtained.\\

\subsection{The case $h_x=0$}  In this case the eigenvalue problem (\ref{sharp eigenvalue equation}) decouples  into two separate problems for the spin-$\uparrow$ and spin-$\downarrow$ components of the wave function, and the magnetic gap energy $\Delta_Z=|h_x|$ vanishes. Accordingly, the eigenvectors (\ref{wpmx>0}) acquire the   simple form
\begin{equation}
w_{-} (K)  \vert_{\Delta_Z=0}=  \left( \begin{array}{c}  1 \\
0 \end{array} \right)  
\,
,
\,
w_{+} (K) \vert_{\Delta_Z=0} =  \left( \begin{array}{c}  0 \\
1 \end{array} \right)
\end{equation}
both for $x>0$ and $x<0$, while the eigenvalues have a quadratic dependence on $K$,   
\begin{equation}
\begin{cases}
E^R_\uparrow(K)=\frac{K^2}{4 E_{SO,R}} -(K+h_z)\qquad &x>0 \\
E^L_\uparrow(K)=\frac{(rK)^2}{4 E_{SO,L}} -(rK+h_z)\qquad &x<0 \\
E^R_\downarrow(K)=\frac{K^2}{4 E_{SO,R}} +(K+h_z)\qquad &x>0 \\
E^L_\downarrow(K)=\frac{(rK)^2}{4 E_{SO,L}} +(rK+h_z)\qquad &x<0  \quad.\\
\end{cases}
\end{equation}
Without loss of generality, we can focus on the spin-$\uparrow$ component of the wave function. The dispersion relation  can be easily inverted 
\begin{equation}
\begin{cases}
K^R_\pm(E)= 2E_{SO,R} \pm \sqrt{(2 E_{SO,R})^2+4E_{SO,R}(h_z+E)}   \\
K^L_\pm(E)= 2 r E_{SO,R} \pm \sqrt{(2 r E_{SO,R})^2+4E_{SO,R}(h_z+E)}   \\
\end{cases}
\end{equation}
In order for  $K^{\nu}_\pm(E)$ to exhibit an imaginary part, one has to consider energies in the range $E<-h_z-E_{SO,\nu}$ and   the most general eigenfunction of energy $E$ can thus be written as 
\begin{equation}
\psi^{(E)}(x)=
\begin{cases}
a \, e^{i\frac{K^R_+(E)}{\alpha_R}x} +b \, e^{i\frac{K^R_-(E)}{\alpha_R}x} \qquad &x>0 \\
c \,  e^{i\frac{K^L_+(E)}{\alpha_R}x} +d \, e^{i\frac{K^L_-(E)}{\alpha_R}x}  \qquad &x<0 \\
\end{cases}
\end{equation}
where $a, \, b, \, c, \, d$ are complex coefficients to be determined. The regularity at $x\rightarrow \pm \infty$ and the continuity in $x=0$ reduce the wavefunction to the form
\begin{equation}
\psi^{(E)}(x)=
\begin{cases}
a \, e^{\frac{iK^R_+(E)x}{\alpha_R}}  \qquad &x>0 \\
a \, e^{\frac{iK^L_-(E)x}{\alpha_R}}  \qquad &x<0 \\
\end{cases}
\end{equation}
while the matching condition (\ref{boundary conditions}) on the first derivative in $x=0$ implies 
\begin{equation}
K^L_-(E)=K^R_+(E)-2E_{SO,R}(1-r)
\end{equation}
whose only possible solution is:
\begin{equation}
\begin{cases}
r^2=1 \\
E=-h_z-E_{SO,R}
\end{cases}
\end{equation}
However, this corresponds to the lowest energy eigenfunction of the continuum, demonstrating that no bound state exists in such case.

\section{Diagonalization strategy  in the presence of a smoothening length} 
\label{AppB}
Here we describe how to  numerically  approach the problem in the presence  of   the RSOC profile (\ref{alpha(x)}) characterized by a finite smoothening length $\lambda_s$,  and when both perpendicular and parallel magnetic field components $h_x,h_z \neq 0$ are present.   To this end,  we impose periodic boundary conditions onto the NW, and express the electron spinor field in terms of  discretized Fourier components $k = 2\pi n/{\Omega}$, namely 
\begin{equation}\label{Psi-Fourier}
\hat{\Psi}(x)= \sum_k  \frac{e^{i k x} }{\sqrt{{\Omega}}}\left( \begin{array}{c} \hat{c}_{k\uparrow} \\ \hat{c}_{k\downarrow} \end{array}\right)  	\quad,
\end{equation}
where $\Omega$ is the (large) NW periodicity  length  and 
$\hat{c}_{k,s}$ denotes the Fourier mode operators for spin $s=\uparrow,\downarrow$. 
The Hamiltonian (\ref{H}) is thus rewritten  in terms of the discretized $k$-basis introduced in Eq.~(\ref{Psi-Fourier}) as  
\begin{eqnarray}\label{H-k-pre}
\hat{\mathcal{H}}  &= &\sum_{k_1,k_2}\sum_{s_1,s_2=\uparrow,\downarrow}  \hat{c}^\dagger_{k_1,s_1} H_{k_1,s_1;k_2 s_2} \,\hat{c}^{}_{k_2,s_2} \quad,
\end{eqnarray}
where
\begin{eqnarray}
H_{k_1,s_1;k_2 s_2} &=& \left[ \left( \varepsilon^0_{k_1}\sigma_0- \mathbf{h} \cdot \boldsymbol{\sigma}\right) \delta_{k_1,k_2} \, - \right. \label{H-k-inhomo} \\
& &\hspace{0.5cm}  -\left. \alpha_{k_1-k_2}\frac{k_1+k_2}{2}\, \sigma_z\right]_{s_1,s_2}\nonumber \quad,
\end{eqnarray}
where $\alpha_q$ is the (discretized) Fourier transform of the RSOC profile $\alpha(x)$.
Specifically,    taking for $\alpha_q$ the following expression 
\begin{equation}\label{alphaq}
\alpha_q =
\begin{cases}
\frac{\alpha_L+\alpha_R}{2} & \mbox{for } \quad $q = 0$ \cr
e^{-\frac{q^2 \lambda_s^2}{32}}
\frac{
\alpha_L
\left(e^{\frac{i q \Omega}{2}} - 1\right)
-
\alpha_R
\left(e^{-\frac{i q \Omega}{2}} - 1\right)
}
{i q \Omega}
& {\rm otherwise}
\end{cases}\ .
\end{equation}
one obtains the (periodic version) of the prototypical profile Eq.(\ref{alpha(x)}) as Fourier series $\alpha(x)=\sum_q \alpha_q e^{i q x}$.

Then, we have performed an exact numerical  diagonalization of the Hamiltonian matrix Eq.(\ref{H-k-inhomo}), thereby obtaining diagonalizing operators $\hat{d}_\xi$ defined through $\hat{c}_a =\sum_\xi U_{a, \xi} \,\hat{d}_\xi$, where $a=(k,s)$ is a compact quantum number notation for the original basis, and $U$ is the matrix of the eigenvectors of Eq.(\ref{H-k-inhomo}). Denoting by $E_\xi$ the eigenvalues, the  NW Hamiltonian can be rewritten as
\begin{equation}
\hat{\mathcal{H}}=\sum_\xi E_\xi\, \hat{d}^\dagger_\xi \hat{d}^{}_\xi
\end{equation}
Finally, to compute the equilibrium expectation values $\langle \ldots \rangle_\circ$ of the operators (\ref{charge-density-def}), (\ref{spin-density-def}), one can re-express  the electron field operator $\Psi_s(x)$ with spin component $s=\uparrow,\downarrow$ in terms of the diagonalizing operators $\hat{d}_\xi$'s,
\begin{equation}
\hat{\Psi}_s(x)=\frac{1}{\sqrt{\Omega}}\sum_{k,\xi} e^{i k x} U_{ks, \xi} \, \hat{d}_\xi
\end{equation}
and to exploit $\langle \hat{d}^\dagger_\xi \hat{d}^{}_{\xi^\prime} \rangle_\circ=\delta_{\xi \xi^\prime} f^\circ(E_\xi)$, with $f^\circ(E)=\left\{1+\exp\left[(E-\mu)/k_B T\right]\right\}^{-1}$ denoting the Fermi distribution function.  For instance, the   density Eq.(\ref{rho-def}) is obtained as
  $\rho(x)=\sum_{\xi} \rho_\xi(x)$, where 
\begin{equation}
 \rho_\xi(x) = \frac{1}{\Omega} \sum_{s=\uparrow,\downarrow} \sum_{k_1, k_2} e^{-i (k_1-k_2) x} \, U^*_{k_1 s,\xi} U^{}_{k_2 s,\xi}\, f^\circ(E_\xi)
\end{equation}
is the contribution arising from the $\xi$-th eigenstate.
In this way, the contribution of each eigenstate (in particular the bound state) can be singled out.



\begin{thebibliography}{99}
\bibitem{nota-su-parity} The case where $\alpha_R<0$ can easily be mapped   into the one  considered here, since the case with a $\alpha(x)$ profile can be mapped into the case $ -\alpha(x)$ by space parity ($p_x\rightarrow -p_x$), as is clear from the Hamiltonian (\ref{H-realspace}).
\bibitem{nota-gauge}  Without magnetic field the problem Eqs.(\ref{H})-(\ref{H-realspace})  can be mapped into the problem of a particle in a quantum well with a potential $E_{SO}(x)= - {m^*}  \alpha^2(x)/2\hbar^2$ by performing the spin-dependent gauge transformation
$\hat{\Psi}(x)=  \exp\left[ i  {m^*}  \sigma_z\int^x_0 \alpha(x^\prime) dx^\prime/{\hbar^2}\right] \, \hat{\Psi}^\prime(x)$.
\end{thebibliography}
\end{document}